\documentclass[a4paper,fleqn]{cas-dc}

\usepackage[authoryear]{natbib}

\begin{document}
\let\WriteBookmarks\relax
\def\floatpagepagefraction{1}
\def\textpagefraction{.001}
\shorttitle{SciServer}
\shortauthors{the SciServer team}

\title [mode = title]{SciServer: a Science Platform for Astronomy and Beyond}                      

\author[1]{Manuchehr Taghizadeh-Popp}
\cormark[1]
\ead{mtaghiza@jhu.edu}
\author[1]{Jai Won Kim}
\author[1]{Gerard Lemson}
\author[1]{Dmitry Medvedev}
\author[1]{M. Jordan Raddick}
\author[1]{Alexander S. Szalay}
\author[1]{Aniruddha R. Thakar}
\author[1]{Joseph Booker}
\author[1]{Camy Chhetri}
\author[1,2]{Laszlo Dobos}
\author[1]{Michael Rippin}
\address[1]{Institute For Data Intensive Engineering and Science, Johns Hopkins University. 3701 San Martin Drive, Baltimore MD 21218, USA}
\address[2]{Department of Physics of Complex Systems, Eotvos Lorand University, Pf. 32, H-1518 Budapest, Hungary}

\begin{abstract}
We present SciServer, a science platform built and supported by the Institute for Data Intensive Engineering and Science at the Johns Hopkins University. SciServer builds upon and extends the SkyServer system of server-side tools that introduced the astronomical community to SQL (Structured Query Language) and has been serving the Sloan Digital Sky Survey catalog data to the public. 
SciServer uses a Docker/VM based architecture to provide interactive and batch mode server-side analysis with scripting languages like Python and R in various environments including Jupyter (notebooks), RStudio and command-line in addition to traditional SQL-based data analysis. Users have access to private file storage as well as personal SQL database space. A flexible resource access control system allows users to share their resources with collaborators, a feature that has also been very useful in classroom environments. All these services, wrapped in a layer of REST APIs, constitute a scalable collaborative data-driven science platform that is attractive to science disciplines beyond astronomy.
\end{abstract}


\maketitle

\section{Introduction}

    The nature of data-driven science today, with the extremely large data sets routinely involved and the collaborative nature of astronomy and many other sciences, creates an urgent need for server-side analysis platforms - a.k.a. \textit{science platforms}. The emerging consensus within the astronomy community\footnote{\href{http://www.stsci.edu/contents/newsletters/2018-volume-35-issue-01/science-platformsserver-side-analytics}{http://www.stsci.edu/contents/newsletters/2018-volume-35-issue-01/science-platformsserver-side-analytics}} is that a science platform should include a server-side suite of data access, analysis and visualization tools, along with the ability to store, upload, download, cross-match and share data with collaborators or students for classroom applications.

    The Sloan Digital Sky Survey (SDSS) science archive \citep{Szalay1999} pioneered the concept of server-side analytics - the backbone of the current science platform concept - in astronomy almost 20 years ago \citep{Szalayetal2000} with the release of SkyServer\footnote{\href{http://skyserver.sdss.org/}{http://skyserver.sdss.org/}} in 2001. SkyServer was the web portal for the SDSS Catalog Archive Server (CAS) and was followed shortly by its spin-off CasJobs\footnote{\href{http://skyserver.sdss.org/CasJobs/}{http://skyserver.sdss.org/CasJobs/}} (CAS + Jobs) in 2003, and both were server-side science portals for SDSS catalog data. Another early SkyServer spin-off was SkyQuery - an astronomy cross-match service \citep{Maliketal2002} that allowed on-demand cross-matching server-side between two or more geographically distributed astronomical archives ba\-sed on a probabilistic cross-match algorithm.

    SkyServer (using the term as shorthand for the suite of services that were accessible from the SkyServer web portal) was the first instance of a publicly accessible astronomical archive available online that embodied the paradigm to "bring the analysis to the data" and allowed users to run queries against very large data sets and manipulate the data server-side in interactive and batch mode. Open SkyQuery, a VO-compliant version of SkyQuery \citep{Budavarietal2004}, was the flagship for the primary mission of the Virtual Observatory (VO), namely to federate the worldwide digital astronomical archives in order to maximize their discovery potential \citep{Szalayetal2002a}.

    Recognizing the foundational role of SkyServer in the astronomical community, the US National Science Foundation (via the Data Infrastructure Building Blocks\footnote{\href{https://www.nsf.gov/pubs/2017/nsf17500/nsf17500.htm}{https://www.nsf.gov/pubs/2017/nsf17500/nsf17500.htm}} program) provided funding to the Institute for Data Intensive Engineering and Science (IDIES) at the Johns Hopkins University in 2013 to extend the server-side analysis functionality beyond catalog data, and to all science domains as well.
    
    SkyServer thus became SciServer \citep{Szalay2018}, a fully featured science platform that took building blocks that had served astronomy for more than a decade, and built a state-of-the-art collaborative and cross-domain science framework on top of them. SciServer is free to use (at its basic level), and anyone can start using it as soon as they register. Data providers and other users with larger resource requirements, such as video cards or nodes for exclusive scientific computing with extra file or database storage, can either host them within SciServer, or deploy an instance of SciServer in their own environment. Although astronomy remains its mainstay (SciServer is the official science platform for SDSS catalog data, for example), SciServer currently supports additional science domains such as turbulence/physics, oceanography, genomics, earth science, materials science and social sciences.

    In this paper, we first give an overview of the SciServer platform with its components and user interfaces. Then we describe our user management and resource access control component. In Section~\ref{Section:DataMangement} we describe the data management components, both in relational databases and the additional data sets accessible on a simple POSIX file system. The most important feature of a science platform is the capability to perform filtering, analysis and visualization close to the data, and we describe SciServer's support for that in Sections \ref{Section:AnalysisCloseToData} and \ref{Section:AnalysisCloseToData:Compute}.
    We describe the usage of the SciServer system so far and the various science use cases we support in Section~\ref{Section:Usage} and close with a summary. An appendix describes the resource management component in some more depth.

    \begin{figure*}
    	\centering
    		\includegraphics[scale=.4]{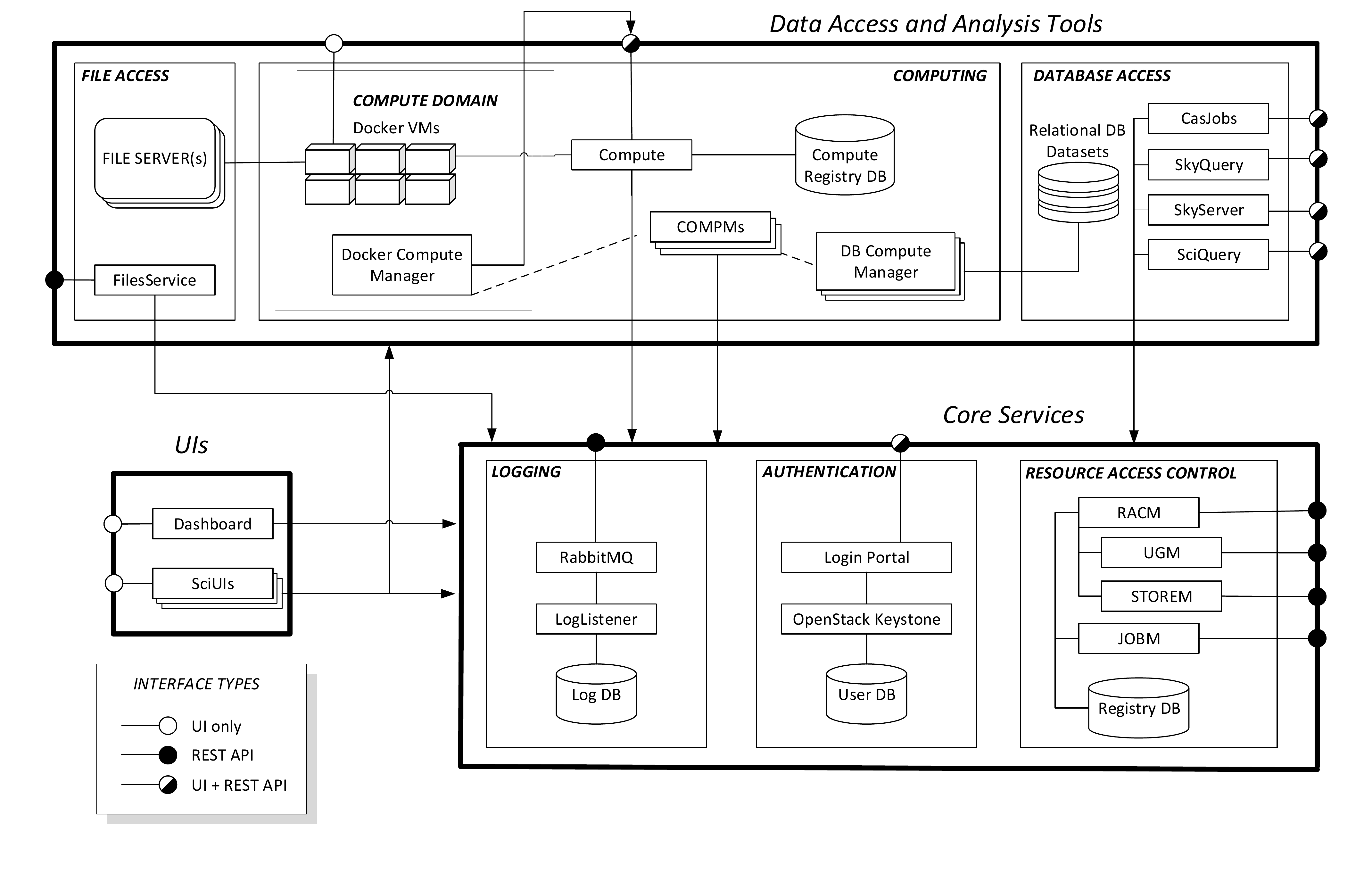}
    	\caption{System architecture in SciServer. The Dashboard web application is the entry point to the system. While the core services in the lower half of the figure are central and unique in a SciServer deployment, the data access and analysis tools in the upper half can be replicated multiple times to scale out storage and computing capabilities. Most components implement REST APIs for communication with other components or external apps. }
    	\label{FIG:Components}
    \end{figure*}

\section{The SciServer Platform}\label{Section:Architecture}

    SciServer consists of a number of core services and applications (authentication, authorization \& resource management, event logging), and a growing set of data access and analysis tools, as shown in Figure \ref{FIG:Components}.
    A key design feature is that most components implement REST APIs for interaction with other components and third-party tools and applications, allowing us to build and extend the system in a scalable, modular way.

    \begin{figure*}
    	\centering
    		\includegraphics[scale=.43]{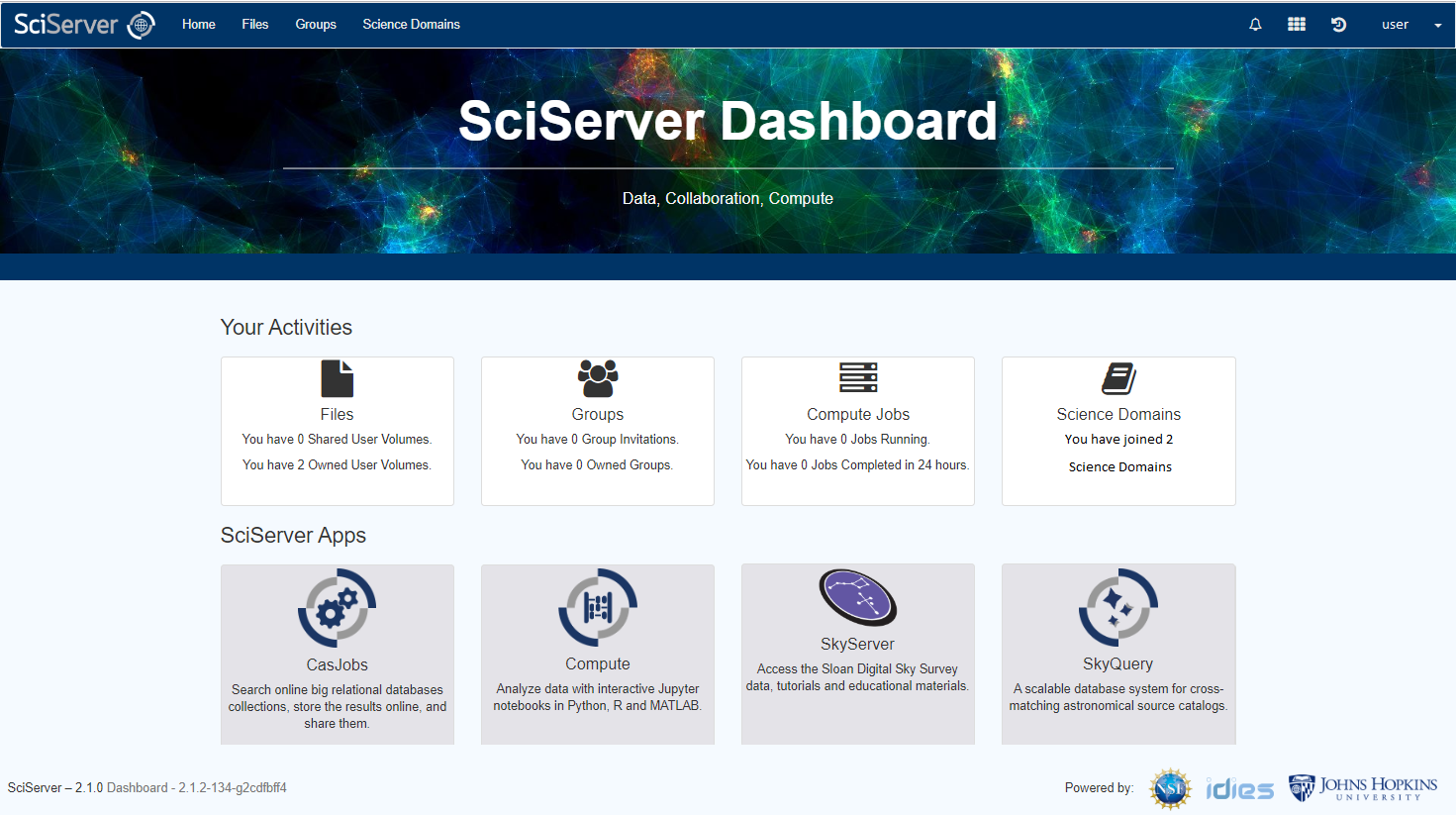}
    	\caption{Screenshot of the SciServer Dashboard. See the text for an explanation of its components.}
    	\label{FIG:Dashboard}
    \end{figure*}

    The core SciServer components in the lower half of Fig.~\ref{FIG:Components} are central to any SciServer deployment. The Authentication module provides a Login Portal web app that allows users to register and log in, and issues session tokens (Section \ref{Section:Authentication}).
    The Authorization and Resource Management module deals with the registration of new resources - such as file volumes, databases, Compute instances, creation of user groups, and the permissions that users and groups have on resources (Section \ref{Section:RACM}). Lastly, the Logging module is in charge of storing user-initiated events or exception stack traces originating from components in the usage history (Section \ref{SubSection:EventLogging}).
    
    The data access and analysis tools in the upper half of Fig.\ref{FIG:Components} can, as opposed to the core components, be replicated multiple times to scale out computing and data storage capabilities. The Database Access module allows users to query big data set tables and store/share their own private databases (Sections \ref{SubSection:Databases} and \ref{Section:AnalysisCloseToData:Compute}); the Files Access module stores data sets on the file system and offers each user a private area for storing/sharing their own files and folders (Sec. \ref{SubSection:FileSystem}). 
    For data analysis, the Compute module allows direct manipulation, exploration, and analysis of the previously-mentioned data sets using py\-thon or R scripts in Jupyter Notebooks \footnote{\href{https://jupyter.org}{https://jupyter.org}} or RStudio \footnote{\href{https://rstudio.com}{https://rstudio.com}} running in Docker containers \footnote{\href{https://www.docker.com}{https://www.docker.com}}, both as interactive sessions and batch job submissions (Sec. \ref{Section:AnalysisCloseToData:Compute}).
    
    External developers can also make use of SciServer REST APIs to build their own web interfaces (SciUIs), which can be integrated with any of the SciServer services, such as data storage and computing tools, for implementing particular science use cases. Examples of these are described in Sec. \ref{SubSection:SciUI}.

    SciServer aims to use Open Source technology as much as possible. Its core services and general-purpose computing framework are developed mostly in Java and run on Linux. However, our historical framework, particularly the astronomy data tools (CasJobs, SkyServer, SkyQuery) was developed on Windows using C\# and Microsoft SQL Server (MS-SQL), so the system as a whole is a hybrid of the two. MS-SQL proved to be well suited for our applications in terms of performance and functionality, though we use MySQL for some of the smaller component-specific registry databases. The science schema for some of our largest databases makes extensive use of MS-SQL’s programmability features, in particular the ability to include compiled C\# code right in the database that is integrated with the system’s CLR (common-language runtime) engine \footnote{\href{https://docs.microsoft.com/en-us/dotnet/standard/clr}{https://docs.microsoft.com/en-us/dotnet/standard/clr}}.
    
    The SciServer deployment of components is in essence designed to be distributed. For example, the core components can be installed on-premise, while the analysis and computing resources could be deployed on virtual machines in the cloud, or on a mix of several local hardware and cloud options. This flexibility enables the building of external "regions" with data and hardware paid, controlled and secured by a particular group of shareholders, and accessed by the public or by a selected group of users.

    Whereas the SciServer instance at JHU is still installed by hand on a cluster of several servers, for new installations we support deployment on Kubernetes clusters \footnote{\href{https://kubernetes.io}{https://kubernetes.io}}. The various web apps run in their own container/pods, with the deployment orchestrated using Helm charts. This greatly facilitates the configuration and scaling-out of the individual components as well as their interoperability.
    
    Although source code for core SciServer components (except for clients, SkyServer, and SkyQuery) is kept private on GitHub repositories  \footnote{\href{https://github.com/sciserver}{https://github.com/sciserver}}, it can be made available for shareholders installing new SciServer instances on their own with a private licensing agreement.

    The SciServer \textbf{Dashboard} (Fig. \ref{FIG:Dashboard}) is the entry point for accessing SciServer applications. Written as a pure Vue.js web client application, it serves as the "Home" screen for a SciServer session. By communicating with the REST APIs of other components, the Dashboard provides the user with a view of all available resources and a way to interact with them, together with a user activity log. The "Activities" section shows the user's current activities and resources, with panels showing the Files, Groups, Compute Jobs and Science Domains activities. Clicking on each panel takes the user to that activity detail page. These pages can also be reached by tabs along the top of the Dashboard screen. 

    On the \textbf{Files} page, the user can create personal folders and upload data sets (see Sec.~\ref{SubSection:FileSystem}). Similarly, on the \textbf{Groups} page, users can create collaborative groups, invite members, and share resources such as data sets with them (Sec. \ref{Subsection:RACM}). 
    The \textbf{Compute Jobs} panel shows the current jobs status and takes the user to the jobs management page (Sec. \ref{Section:BatchJobsInCompute}).
    The \textbf{Science Domains} tab shows the domains hosted in SciServer offering access to public resources that may be only of interest to members of a particular science domain (see Section~\ref{SubSection:ScienceSupport}). 

    Below the Activities layer, the Dashboard page provides links to the SciServer Apps: \textbf{CasJobs}, \textbf{Compute}, \textbf{SkyServer}. and \textbf{SkyQuery}. These are described in detail under "Analysis Close to the Data" in Sections \ref{Section:AnalysisCloseToData} and \ref{Section:AnalysisCloseToData:Compute}.

\section{User management, Resource Access Controls and Sharing}\label{Section:RACM}

    SciServer enables complex scientific analysis on Pe\-ta\-byte-scale data sets, some of which are only accessible to a restricted group of users. Moreover, many of the analysis jobs are too large to be executed in interactive mode, and their results must be stored on the system for further analysis and visualization. SciServer supports this by providing users with personal workspaces that by default are only accessible to their owners, but since science is inherently collaborative, SciServer also enables the sharing of workspaces by geographically separated science teams. 
    
    To support these collaborative projects, SciServer has implemented a very flexible user authentication and authorization mechanism that will be described in the following subsections.

    \subsection{User registration, authentication and single-sign-on}\label{Section:Authentication}

\begin{figure}
    	\centering
    		\includegraphics[scale=.50]{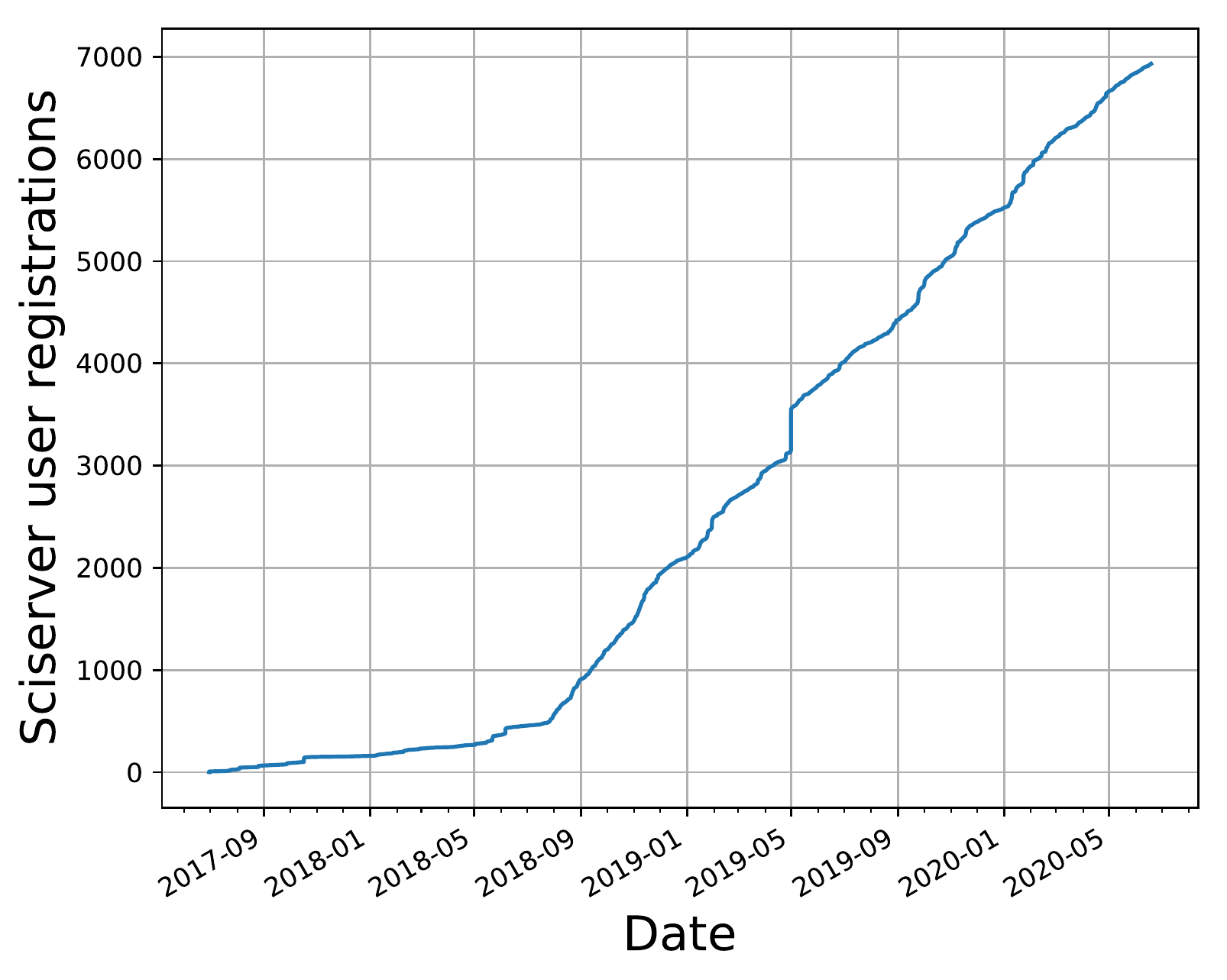}
    	\caption{Cumulative number of user registrations with SciServer as a function of time. Note the increase in number of registrations since the July 25th, 2018 release of SciServer featuring a new Dashboard, compute batch jobs, and resource access control management. The jump of user registrations in May 2019 is due to the migration of old CasJobs users and groups into SciServer.}
    	\label{FIG:LoginPortalEvents}
    \end{figure}

    New users can register in the SciServer Login Portal\footnote{\href{https://apps.SciServer.org/login-portal}{https://apps.SciServer.org/login-portal}}, where they need to provide a unique username and email address, and enter a secure password. 
        Registration has so far been open to all external users, and Fig.~\ref{FIG:LoginPortalEvents} shows the cumulative growth of our user base since mid-2017. Note that this excludes all the users that had been registered previously in CasJobs. Some jumps indicate classes or schools, with a larger number of students registering almost simultaneously. The big jump in April 2019 was due to an internal operation that force-migrated a subset of users from the old CasJobs system to our current system.
    The increase starting mid-2018 is due to the first public release we  made of the full SciServer functionality, in particular its collaborative features. 

    Authentication in all SciServer components is handled by a common identity service. We use OpenStack Keystone \footnote{\href{http://docs.openstack.org/keystone}{http://docs.openstack.org/keystone}} for that purpose. It provides a simple and convenient way of token-based authentication and some high-level authorization. When a user logs in, a new token is generated with a lifetime of 24 hours. The different components in SciServer use the token to identify the user and pass it along when communicating with other components through their REST APIs.
    
    Interactive sign-on is also possible with external identity providers. In that scenario, we use Keycloak \footnote{\href{http://www.keycloak.org}{http://www.keycloak.org}} as identity broker for initial sign-on, however at the current stage the Keycloak user ID is mapped to a Keystone user ID, and Keystone tokens are used for all subsequent operations. Currently our Keycloak service is configured to use Globus Auth \footnote{\href{https://docs.globus.org/api/auth/}{https://docs.globus.org/api/auth/}}, which in turn can serve as a broker for other identity providers, including Google, ORCID, and many academic institutions. Keycloak could also be configured to work with other external identity providers directly.

    \subsection{Resources Access Control and Groups}\label{Subsection:RACM}
    The various SciServer components illustrated in Figure~\ref{FIG:Components} manage a variety of types of \textbf{resources} - the abstraction by which we refer to all data and computational entities to which users can be given access. Examples are relational databases managed by CasJobs, data sets in "flat files" managed by the Files Service, and also specific computational resources managed by SciServer Compute.
     The access to these resources is controlled by SciServer's Resource Access Control Management (\textbf{RACM}) component. This component provides formal definitions of the types of resources and the application components that manage them, as well as the actions that can be performed on them. RACM also stores representations of the actual resources along with instances of these \textbf{resource types}, and assigns privileges to users and/or user groups to perform certain actions on them.
    
    In RACM, the SciServer applications such as the Files Service or Compute are represented explicitly as \textbf{resource contexts}, and they use the RACM APIs for storing access privilege assignments on the resources they manage, such as data sets or Compute environments. They use these APIs also for checking the permissions a user has on their resources, i.e. whether, at any given point in time, a given user is allowed to undertake a given action on a given resource.

    This abstract description of the resources in SciServer is represented as a formal object model that, together with its implementation on a MS-SQL relational database, is described in some more detail in Appendix~\ref{Appendix:RACM} and is illustrated there in Fig.~\ref{FIG:ClassDiagramRACM}.
    
    RACM has specialized REST API endpoints for user and group management (UGM), file services and storage management (STOREM), and for batch job submission (JOBM). In addition, its abstract API can also be used directly to define and manage new types of components and their resources. This, for example, is how CasJobs defines its database contexts and their access rights.
    
    In addition to being able to give, or restrict, access to large data sets and compute resources, SciServer also supports sharing of resources created by users themselves. Users can define "User Volumes" on the file system (see Sec. \ref{SubSection:FileSystem}) and share these with others, generally to support some collaborative projects.
    To simplify this SciServer allows users to create \textbf{user groups} in the Groups tab in the Dashboard. Groups have a name and a description and, once created, the owner of a group can invite other users to the group, who can accept or decline this invitation. When inviting a new member, the creator can assign to them an ADMIN or MEMBER role, the former providing permissions to invite, remove or promote other users in turn. In principle, Groups can be nested, but this is not yet implemented in the current version of SciServer.
    Resources can now be shared with the group, and members inherit these privileges. Upon registration, all users are automatically added to the \textbf{Public} group, which offers access to all public resources of SciServer, though ordinary users are not allowed to share resources with this group.

\section{Data management}\label{Section:DataMangement}

SciServer offers a plethora of public and private data sets, either in the form of relational databases or simple flat files on a file system. This section describes how they are managed within the system, and Table \ref{Table:ListOfDatasets} shows a list of some of them. More detailed descriptions can be found on the SciServer\footnote{\href{http://www.sciserver.org/datasets/}{http://www.sciserver.org/datasets/}} and SkyQuery\footnote{\href{http://www.voservices.net/skyquery/Apps/Schema/Default.aspx}{http://www.voservices.net/skyquery/Apps/Schema/Default.aspx}} websites.    

\begin{table*}[width=1.7\linewidth,cols=5,pos=h]
\caption{List of major data sets from several science domains publicly available in SciServer, together with their types, main SciServer services for accessing them, and access types. Note that SkyQuery data sets contain mostly sky positions and might not feature the full released data content.}\label{Table:ListOfDatasets}
\begin{tabular*}{400pt}{LLLRL}
\toprule
Science Domain & Data sets & Type & Size\footnotemark & Service \\
\midrule
Astronomy & SDSS DR1-15, Stripe82, RunsDB, & database & 150TB & CasJobs, SkyQuery   \\
          & First, Gaia DR2, Galex GR6, &   &  &    \\
          & 2DF, 2MASS, Rosat,  &   &  &   \\
          & ACVS, AGC, AKARI, CHANDRA,  & database  & 100TB & SkyQuery   \\
          & CNOC2, COMBO17, CS DR2,  &    &  &    \\
          & DEEP2 DR4, DES, DLS, FORS, &    &  &   \\
          & FUSE, GAIA DR1, GDDS, HDF,  &    &  &    \\
          & HERSCHEL, IRAS, K20, KEPLER,  &    &  &    \\
          & KPGRS, LAMOST DR1, LBG3Z,  &    &  &    \\
          & MGC, NDWFS, NVSS, OGLE III,   &    &  &    \\
          & PSCz, RC3, SPITZER, SSRS,  &    &  &    \\
          & TKRS, USNOB, UZC, VIPERSDR1-2, &    &  &    \\
          & VLA, VVDS, WiggleZ, WISE, &    &   &    \\
          & WMAP, zCOSMOS &    &   &   \\
Cosmology & Millennium Run Databases  & database & 40TB  &  CasJobs  \\
Astronomy & SDSS DR7 DAS, MaNGA,  & files &  100TB &  Compute  \\
          & SDSS DR14/16 Spectra,   &  &    \\
          & Heasarc, Kepler, WFIRST &    &  &    \\
Cosmology & Millennium Run Raw Data, & files & $\gtrsim$1PB & files  \\
          & Indra Simulations,  &  &  &      \\
          & Eagle Simulations  &  &    &    \\
Materials Sci. & FragData & database & 10TB  &  CasJobs  \\
Genomics & Recount2 & files & 13TB  &  Compute  \\
Oceanography & various simulations & files & 17TB  &  Compute  \\
Turbulence & various simulations & database+files & 450TB  &  Compute  \\
\bottomrule
\end{tabular*}
\end{table*}
\footnotetext{Sizes of data sets are for a single instance, typically there are 2-3 instances of each for fault tolerance and load balancing.}

    \subsection{Databases}\label{SubSection:Databases}
    
    When SDSS introduced SkyServer in 2001, and followed soon thereafter by CasJobs \citep{OMullaneetal2005} in 2003, it was the first instance of a large astronomical data set being hosted and served online from a commercial database management systems  \citep{Thakaretal2008}. Since then, the SkyServer system has executed nearly half a billion SQL queries \footnote{\href{http://skyserver.sdss.org/log/en/traffic/}{http://skyserver.sdss.org/log/en/traffic/}}. The astronomical community's adoption of SQL and databases over the first ten years of SkyServer was documented and analyzed in two 2014 reports \citep{TenYrs1Raddicketal2014, TenYrs2Raddicketal2014}.
    While they may not be the best solution for every type of Big Data workload (e.g., Google has specialized solutions for its online search engine), databases excel at certain tasks that are common to astronomical data analysis:
\begin{itemize}
    \item \textbf{Data Integrity}: databases excel at maintaining and assuring the accuracy and consistency of large data sets. Primary keys that enforce uniqueness constraints, and foreign keys and constraints that define relationships and enable the validity of data to be checked, are invaluable in maintaining data integrity and consistency in large and complex data sets. In SDSS, these have enabled us to find problems upstream of the science archive - in the raw data pipeline processing and resolve steps - on more than one occasion.
    \item \textbf{Optimized Data Access}: databases provide a common  lan\-guage (SQL) for optimized access to large and complex data sets across a wide range of query workloads. Clustered and non-clustered indices allow da\-ta\-base designers to optimize data access for a virtually unlimited set of use cases and query patterns. Performance tuning tools built into the database further facilitate this task.
    \item \textbf{Server-side Analytics}: modern database platforms - with features like stored procedures and user-defined functions - allow complex science logic to be encoded in the database, right next to the data, thereby minimizing data movement. We refer to this as the "science schema".
\end{itemize}
When choosing a database platform, it is important to prioritize the technology and features that are likely to ensure years of reliable and responsive data management, such as mature technologies, major vendors responsive to market trends, performance, support for server-side scientific data analysis, ease of administration, etc.  SDSS chose MS-SQL - a relational database management system - as the database platform because of its performance (highly rated query optimizer), pricing, vendor support, excellent native floating point support, and the ability to encode science schema into the database via stored procedures and functions \citep{ThakarSzalay2003}.\\ Subsequently, MS-SQL also provided integrated support for compiled code in the database via its Common Language Runtime (CLR) feature, which made the science functions encoded in the database run orders of magnitude faster than before. Examples of that are the HTM and Spherical libraries for spatial indexing and sky survey footprint support \citep{BudavariEtAl2010}, and the CFunBASE library for computing cosmological distances and times  \citep{Taghizadeh-Popp2010}. One issue with MS-SQL was that it originally only ran on Microsoft Windows operating systems, but that has changed in the past few years and it is now fully available on Linux as well.

Support for MS-SQL is grandfathered into SciServer, but it is meant to be a platform-agnostic system, and currently supports Postgres and MySQL database platforms via the SciQuery component described in Sec. \ref{SECTION:Sciquery}.

    \subsection{File-based Data}\label{SubSection:FileSystem}

        In addition to the many catalog data sets that SciServer hosts, mostly as relational databases, there are many file-based data sets hosted on a POSIX file system under special folders (directories) called \textbf{data volumes}. 
        Users can directly access the data volumes from within the SciServer Compute app (by means of NFS mounts in the compute VMs of paths in remote storage servers) for further data analysis and exploration.
        
        Astronomical file-based data sets include the SDSS DR7 DAS\footnote{\href{http://das.sdss.org/}{http://das.sdss.org/}})  with FITS images, a local copy of the SDSS SAS (Science Archive Server\footnote{\href{http://data.sdss.org/sas/}{http://data.sdss.org/sas/}}) that includes all the FITS files corresponding to the SDSS imaging and spectra since DR9, MaNGA datacubes, etc. SciServer hosts many file-based data sets from other science domains, including Genomics, Materials Science, Oceanography, Earth Science, Social Sciences, etc. (see Table \ref{Table:ListOfDatasets}). 
        These data volumes can be made public, or shared between a private collaboration, using the Groups feature.
        
        Users can also create, upload, store, or share data in their own folders called \textbf{user volumes}, 
        which are in turn located under \textbf{root volumes} on an XFS file system in remote storage servers.
        This data path, as seen from inside the Docker containers and exposed by the Jupyter or RStudio applications in SciServer Compute, is .../\verb+RootVolume+/\verb+UserName+/\verb+UserVolume+/. \\
        All user volumes created under the \verb+Storage+ root volume are backed up and subject to a quota of 10GB of data in total. The quota for each new user is enforced at the moment of volume creation by using specific XFS commands. 
        In charge of that is the SciServer Quota Manger Java Spring Boot application that runs in the remote XFS storage servers. 
        On the other hand, all user volumes created under the \verb+Temporary+ root volume are not subject to quotas, and can be used as scratch space, although they are not backed up. One default user volume is created automatically under each root volume, each time a new user registers and enters the Dashboard application for the first time. 
        
        The \textbf{Files Service} is a modular Java Spring application that implements a REST API, and is in charge of managing file system operations by means of HTTP requests from the user. Users can use it to remotely upload, delete, or create files or folders, as well as for creating and sharing user volumes (in read-only or read/write mode). Mainly, these requests are initiated and executed in the Dashboard application.
        All incoming requests to the Files Service are first authenticated and permissions checked against RACM's STOREM API (Sec. \ref{Section:RACM}). Users can share user volumes, but not individual files or sub-folders. It is too easy to create or delete from within the SciServer Compute application, and without more advanced event tracking on the file system it is difficult to keep the state of the file system in sync with the entries registered in the RACM database.

\section{Analysis Close to the Data: SQL}\label{Section:AnalysisCloseToData:SQL}
\label{Section:AnalysisCloseToData}
    SciServer incorporates the "bringing the analysis to the data" paradigm throughout its design. Since it hosts both relational databases and file-based data sets, SciServer offers a variety of ways that users can do their analyses close to the data. For databases, the most direct and efficient way to retrieve the data is with SQL, and several SciServer tools allow users to submit SQL queries, either directly or indirectly, to the database servers in the most efficient manner. For file-based data sets, there are other ways to query and retrieve the data while minimizing data movement.  Here we describe how the major tools enable SciServer users to do their server-side analytics close to the data.

    One of the pivotal sociological changes that the era of big surveys has brought about is that astronomers have, for the most part, learned how to talk to databases: they have taught themselves how to program in SQL. The International Virtual Observatory Alliance (IVOA) adopted ADQL (Astronomical Data Query Language) as a standard, which was a subset of SQL \footnote{\href{http://www.ivoa.net/documents/latest/ADQL.html}{http://www.ivoa.net/documents/latest/ADQL.html}}. IVOA's TAP (Table Access Protocol) standard recognizes ADQL and also allows for native SQL to be passed through to the underlying services \footnote{\href{http://www.ivoa.net/documents/TAP/}{http://www.ivoa.net/documents/TAP/}}. 
    
    SQL is the most powerful way to formulate queries and build workflows for a relational database, and hence it is the most powerful tool for server-side analysis with databases. SciServer fully supports the submission of SQL queries to the data sets that are hosted. There are other ways to submit queries, including web forms, visual tools and cross-match tools, but ultimately every query that is handled by the database query engine is received as a SQL query.

    Depending on the scope of a query, it can finish executing in a few seconds or a few hours, or even a few days for the most intensive queries on the largest data sets. Our analysis of SkyServer query performance showed early on that it was very important to segregate different types of query workloads to different servers. In particular, the same instance of the query engine should not handle both quick (interactive) and long (non-interactive) queries. Generally speaking, quick queries are handled in a synchronous manner that is suitable for a browser based web portal. Queries that take longer than a few minutes must necessarily be handled in asynchronous or batch mode.
    
    The SciServer services that handle SQL queries against the hosted catalog data sets are SkyServer, Cas\-Jobs, SkyQuery and SciQuery.  We describe these below.  
    
    \subsection{SkyServer}
    SkyServer \citep{Szalayetal2002} is a synchronous, interactive, browser-based web portal that allows users to submit limited queries against the SDSS data sets hosted in SciServer.  The queries supported by SkyServer are limited in both space and time: there are limits to the output size (up to a maximum of 500,000 rows) and query execution time (10 minutes) built into the web application. As mentioned above, SkyServer predates SciServer and is in fact the original web portal that offered SDSS data to the world. It has undergone several changes through the years including technology upgrades and interface redesigns.  In SciServer, the back end of SkyServer has been re-engineered as a web service with a REST API. It has also been better integrated into the SciServer framework, with an optional login now available with the SciServer credentials, so the user's history of all their SkyServer activity can be stored and accessed.  Also, they can save results of SkyServer (synchronous) queries to their MyDB if they are logged in.
    
    SkyServer remains the primary public portal for SDSS catalog data worldwide, and is indeed still being widely used, receiving half a billion web hits per year\footnote{\href{http://skyserver.sdss.org/log/en/traffic/}{http://skyserver.sdss.org/log/en/traffic/}}. The SkyServer user interface last received a major upgrade in 2008, and it is currently in the process of being upgraded again, to bring it up to the current web-UI standards and technologies. A new version of SkyServer will be released early in 2021. Source code is available to the public in GitHub repositories 
    \footnote{\href{https://github.com/sciserver/skyserver-DR16}{https://github.com/sciserver/skyserver-DR16}}\footnote{\href{https://github.com/sciserver/skyserver-ws}{https://github.com/sciserver/skyserver-ws}}.
    
    \subsection{CasJobs}
    The wide range in query workloads and the need to segregate them, along with the imperative to minimize data movement, were the primary driving factors that led to the creation of CasJobs - the batch query workbench service that we developed \citep{LiThakar2008} in addition to the synchronous, browser-based SkyServer web portal for SDSS catalog data. CasJobs allowed users to submit queries that are queued for execution in a batch queue and typically run over several hours. This of course necessitated the storing of query results server-side so that the user could download or otherwise use them at a later time.
    
    The solution we came up for storing query results server-side was to give every user their own personal SQL database - called \textbf{MyDB}.  Users can not only store, reuse and share database tables in their MyDB, but they have access to all the features that a database engine like MS-SQL provides - they can write their own custom stored procedures and functions, create complex SQL workflows that include variables and temporary tables, and enhance performance and usability by creating their own indices and views (virtual tables) against large tables. We also gave users the option to create special "public" MyDBs to share results of research projects or papers with the community (an example is the DeepPM data set for the deep proper motions data set from  \cite{Munnetal2014}).
    
\begin{figure}
	\centering
		\includegraphics[scale=.25]{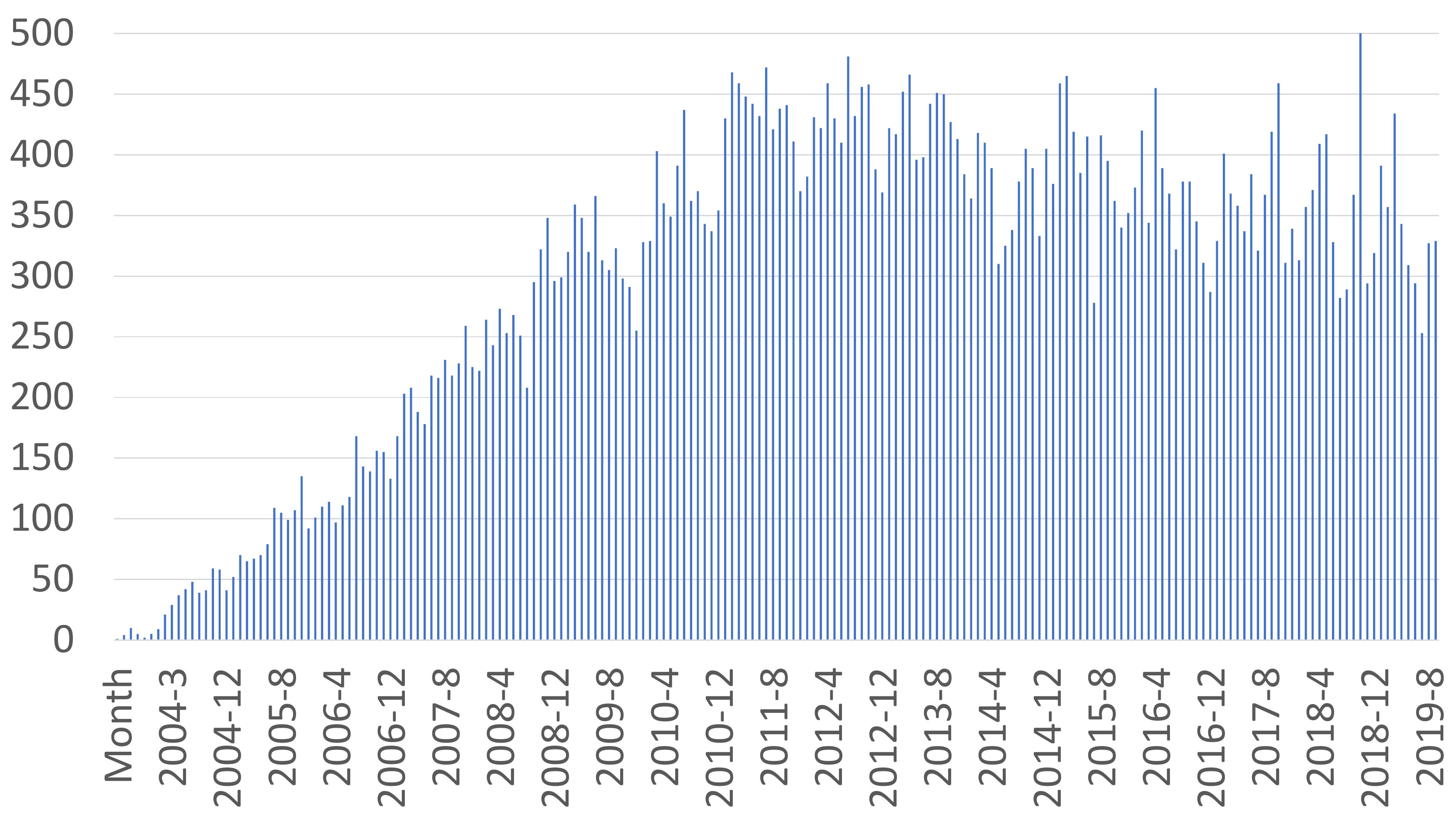}
	\caption{The number of unique users per month running queries in CasJobs, since its inception in 2003.}
	\label{FIG:cjusers}
\end{figure}

    The private SQL/database workbench server-side at the disposal of the user became extremely popular, and within a few years CasJobs had thousands of user accounts from users worldwide.  Given that there are $\sim$ 10000 astronomers worldwide, this was a significant fraction of the world astronomical community. There are currently more than 14,000 CasJobs users, of which over 11,000 have accessed their MyDBs in the past 5 years. The active users per month since the beginning are shown in Figure \ref{FIG:cjusers}.
    
    The CasJobs \textbf{Query} page supports SQL queries in two modes - synchronous or \textit{quick mode} queries, and a\-syn\-chro\-nous or \textit{batch mode} queries. The output from both types of queries can be saved in your MyDB. This option is mandatory for batch mode, for obvious reasons.  If you forget to specify a MyDB table destination for a batch query, CasJobs will add one for you. All SkyServer SQL queries are guaranteed to work on CasJobs.  CasJobs adds a few syntactical elements of its own in order to enable the server-side query and analysis functions. 

    CasJobs has several features that make it a very versatile SQL-based analysis facility.  Unlike SkyServer,  it supports the concept of a "session" so you can use variables and temporary tables that are valid for the duration of your session. Since CasJobs hosts multiple databases from various projects including SDSS, Gaia, GALEX and several other astronomical data sets, it introduces the concept of a "context", which is the current data set that you are submitting your query to. So there are multiple contexts available for the different SDSS data releases (DR1 through the latest, DR16). There is also a context corresponding to your MyDB.

    There is a \textbf{History} page that saves every single query that you ever submitted to CasJobs, along with a search and filter capability.  You can select a previous query and rerun it, with or without modifications. There is an \textbf{Import} page that allows you to upload your own data to CasJobs. A schema browser (via the MyDB page) allows you to browse all the tables, views, functions and stored procedure in all the contexts available to you. You can even browse the SQL code for functions and procedures.

    CasJobs has a \textbf{Groups} page that allows users to share their own MyDB tables with collaborators. This Groups functionality is integrated with the Groups feature in SciServer (see section \ref{Subsection:RACM}), so your MyDB tables will appear on the SciServer Groups page along with other resources     that you have shared with your group members.

    CasJobs login is now integrated with the SciServer single sign-on system, so there is no separate CasJobs account any more.  Every time a new user registers with SciServer, they get access to CasJobs and their own MyDB.  CasJobs is a web application that talks to a web service layer via a REST API. CasJobs API calls are integrated into other SciServer components like SciScript and Compute, via the CasJobs Python and R library, So you can interact with CasJobs from a Python or R script (e.g., in a Jupyter notebook or SciScript).
    
    Although the initial default MyDB space is kept very low intentionally (0.5 GB), so that users new to SQL don't accidentally fill up their MyDB (with an incorrectly formulated JOIN clause, for example), MyDB space is increased upon request quite liberally up to tens of GB.  CasJobs also now gives users the ability to store arbitrarily large query results in a temporary, scratch database called \textbf{MyScratch}. The tables stored in MyScratch are cleaned up periodically, and do not count towards the MyDB space limit.
    
    \subsection{SkyQuery}\label{Subsection:SkyQuery}
    
    SkyQuery is a SciServer application that allows users to perform massive cross-match operations on astronomical catalogues. The version of SkyQuery that is integrated with SciServer is its third incarnation \footnote{\href{https://github.com/sciserver/skyquery-all}{https://github.com/sciserver/skyquery-all}}, following the initial SkyQuery prototype that was released in 2002, and the VO-compliant Open SkyQuery service that was part of the VOServices site hosted at JHU from 2004 to 2014. The current site \citep{Budavarietal2013} is a re-engineered version that has been integrated with the SciServer single sign-on system and the user activity logging system (section \ref{SubSection:EventLogging}), and it includes the option to save the results of the cross-match query to the user’s MyDB or MyScratch database contexts in CasJobs. Apart from the website, SkyQuery functionality can be accessed by 
    a Python client library, which can be executed from Jupyter notebooks in SciServer Compute (see sections \ref{SubsubSection:Compute} and \ref{SubSection:SciScript}).
    
    SkyQuery implements an N-way Bayesian probabilistic join that uses the Zones algorithm \citep{Gray2006} to speed up the spatial search aspect of the cross-matching. The entire cross-match algorithm is implemented in SQL with a few user defined functions written in C\#, to benefit from the optimizer built into the query engine \citep{Budavarietal2013}. SciServer hosts all required data sets on a dedicated cluster; this co-location allows for much better performance. SkyQuery is built for high-throughput parallel hardware, and a fully parallel and asynchronous job system and I/O libraries were developed to orchestrate parallel query execution across a cluster. The details on the architecture and design of SkyQuery are beyond the scope of the present paper, but are available in \cite{Budavarietal2013}.

    \subsection{SciQuery}\label{SECTION:Sciquery}
    
    The ability to submit SQL queries to any database platform would obviously be very useful for heterogeneous data sets that are hosted on different platforms.  CasJobs only supports MS-SQL databases. Even though the full MS-SQL functionality is now available on Linux, there are many astronomical data sets that are hosted on other database platforms, and it is in general not feasible to migrate these to MS-SQL. Furthermore, CasJobs is still implemented in C\# and ASP.Net and requires a Windows IIS web server for its deployment, unlike the other SciServer components, which run on Java. 
    
    To resolve these issues we have started building a new application, \textbf{SciQuery}, that will replace CasJobs, and is written in Java.  SciQuery is a platform-agnostic service for running SQL queries against relational databases. It is able to launch asynchronous queries not only to MS-SQL, but also to PostgreSQL and MySQL. Currently hosted astronomy data sets that are accessible via SciQuery are the SDSS collaboration's MaNGA data set and the SuMIRe collaboration's HSC-PFS data set. For each of these, the data is hosted as PostgreSQL databases operating in read-only mode.
        
    SciQuery uses the same mechanism for running batch jobs that is used by Compute for submitting asynchronous Docker jobs, as described in Section~\ref{Section:BatchJobsInCompute}. Here, the Database-COMPM service picks the query Job definition from the job queue and submits it directly to the database server.  The tabular result sets are then written to the user's Jobs directory under a user volume, or sent to the user's MyDB database.

    \section{Analysis Close to the Data: Compute}\label{Section:AnalysisCloseToData:Compute}
    Many users of the CasJobs and SkyServer services desired to have more advanced tools available to analyze and visualize the query results server-side. With the increasing size of the data, downloading even filtered results was becoming impractical. Also, they were interested in joining the catalogue data to original data sets such as the images and spectra obtained by SDSS, which were capabilities not available through either the SQL back-end or the SkyServer system.

Hence it was one of the important goals of SciServer to support such more advanced, server-side analysis, and SciServer {\em Compute} is its implementation.
SciServer {\em Compute} allows users to write and execute advanced data analysis and exploration code in Python, R, Julia and a number of other computer languages server-side, generally by running a Ju\-pyter\footnote{\href{https://jupyter.org}{https://jupyter.org}} notebook server in their web browser. The following sections describe this SciServer component  in more detail.

    \subsection{Containerization and Scalability}

SciServer Compute provides users with their own virtual, computational environment, running server-side on a designated VM cluster. 
These environments are implemented using the Docker\footnote{\href{https://www.docker.com}{https://www.docker.com}} container technology. The containers isolate the execution of programs and services from the underlying operating system. And they can be configured to only have particular data and user volumes available for seemingly local access.
        This "containerization", or operating-system-level virtualization, offers an elegant way to create an isolated environment for each user
        and has become the standard way by which astronomical science platforms provide compute capabilities to their users.
        
        Compute allows users to create one or more containers, and manages their life cycle. Users create a container by choosing one from  a growing list of pre-defined Docker {\em images}. These images define the computational environment and are created from Docker files\footnote{\href{https://docs.docker.com/engine/reference/builder/}{https://docs.docker.com/engine/reference/builder/}}. Typically in SciServer, the image installs Jupyter in the container, together with any particular software libraries and user/data volumes  needed for a particular science domain workflow. When the user starts a container, Compute will start the Jupyter server inside the container, and create a proxy URL for the server that is accessible only to the user. At the moment, a maximum of 3 containers can be active for a user at any given time. The system will stop containers (i.e., persist them to disk) after 3 days of user inactivity in order to preserve RAM and CPU resources on the host nodes.
        
        The container orchestration implemented in Compute is similar to, but predates, the popular JupyterHub\footnote{\href{https://jupyter.org/hub}{https://jupyter.org/hub}} tool. SciServer Compute adds features to its orchestration that are not (yet) available in that tool. In particular, Compute connects to RACM when a user makes a request to open a container, and evaluates whether the requested images and data or user volumes to be mounted are accessible to the user.

        Compute organizes its underlying hardware and software resources in so-called \textbf{Compute Domains} (see Fig.~\ref{FIG:Components}). A Compute Domain represents a collection of compute nodes, possibly Virtual Machines (VMs), that can run Docker containers. All nodes in a compute domain have access to the same set of Docker images and can mount the same data and user volumes. 
        In its current incarnation, Compute creates a container in a  domain by choosing one of its nodes in a round-robin-fashion and calling the Docker API to spawn the new container \citep{Medvedev:2016:SCB:2949689.2949700}. 
        Compute's registry database stores the configuration metadata for all running containers. 

        In the current release of SciServer, when more computing resources are needed, compute domains can be scaled out only by adding more nodes or VMs and configuring them for running compute containers manually.
         We are in the process of migrating Compute to use Kubernetes\footnote{\href{https://kubernetes.io}{https://kubernetes.io}} for container orchestration. This will greatly simplify scale-out and will provide enhanced robustness against node or container failure.
         It will also facilitate the creation, sharing and reuse of custom images from Docker containers that have been modified by users and will allow us to implement dynamically mounting new data volumes in running containers more easily.
        
        The SciServer Compute ecosystem can be expanded by deploying new Compute Domains on external systems, for example in an external data center, or on one of the commercial clouds, while registering them in the central SciServer Compute registry so they can be managed using the same interfaces.
        The advantage of this pattern is that, while being integrated with the SciServer user management, resource control and data access layers, the owner has full control over access to data and compute resources local to the domain, as well as a direct control on the operational costs (i.e, a domain in a commercial cloud could be shutdown when not needed). Incorporating such support in SciServer is currently under active development.

\begin{table*}[width=1.98\linewidth,cols=5,pos=h]
\caption{List of some Docker images available in SciServer Compute.}\label{Table:ListOfDockerImages}
\begin{tabular*}{470pt}{LLL}
\toprule
App & Images & Purpose  \\
\midrule
Jupyter Notebooks/Lab   & Astroinformatics 2018          & with tools for Astroinformatics summer school    \\
                        & BeakerX          & kernels and extensions to Jupyter with JVM languages \\
                        & CASA          & with Common Astronomy Software Application  \\
                        & Census, Geo      & with GIS tools                                       \\
                        & Heasarc          & with Common Heasoft packages  \\
                        & JH Turbulence DB & with libraries for accessing Turbulence database     \\
                        & Julia            & includes Julia kernel                                 \\
                        & LSST             & with LSST science pipelines and packages             \\          
                        & Machine Learning          & with TensorFlow, PyTorch and Keras  \\
                        & Marvin           & with Marvin tools for analysing MaNGA data set        \\
                        & Montage          & with MontagePy for mosaicking and visualizing astronomy images  \\
                        & Oceanography     & with tools for analyzing ocean circulation simulations             \\
                        & Python + R       & general purpose base image with development tools    \\
                        & PyTorch          & with PyTorch deep learning library                             \\
                        & Recount          & with Bioinformatics tools for Recount2 data analysis  \\
                        & SoFIA          & with SoFIA tools for finding HI sources   \\
                        & WFIRST          & with WFIRST software tools \\
                      
RStudio                 & Python + R       & general purpose base image with development tools     \\
            
\bottomrule
\end{tabular*}
\end{table*}

    \subsection{Interactive Compute}\label{SubsubSection:Compute}
        An interactive Compute session typically starts when, in the Compute UI, a user creates a container from a Docker image that starts a Jupyter server. The system currently also offers RStudio, and could support other applications implementing a web UI for the Docker environment. Users are presented with the traditional Jupyter interface from which they can start new notebooks or open existing ones. The container will have the user's private volumes mounted and they can upload or download files.
        Most data collections hosted by SciServer as Data Volumes can be mounted inside the Docker containers, and analyzed directly in the Notebooks. 
        
        The libraries that can be used in a notebook depend on the image that the user chose when they created the container. The system offers a collection of public and private Docker images, customized for different workflows, or science domains, see Table \ref{Table:ListOfDockerImages}. The images are based on a base 
        Linux image, on top of which we create several layers adding new specific software features. 
        Most images use a generic "Python+R" image, which contains a set of most commonly-used libraries, including astropy\footnote{\href{https://www.astropy.org}{https://www.astropy.org}}. This image has been specialized further for example  with tools for machine learning (e.g. the "PyTorch" and "Machine Learning" images), or special astronomical libraries such as the "Montage" and "LSST" images
        
        The Jupyter interface also offers a terminal service. This allows users even more flexible control over their environment. For example they can install libraries using the standard "pip" or "conda" installers, that are not available in their container; they can "git clone" software shared on GitHub and they have access to most of the gcc-family of compilers for building legacy applications.

    \subsection{SciScript}\label{SubSection:SciScript}
        We have written special-purpose libraries in Python and R that allow users to access the various SciServer components from within a compute container.
        {\em SciScript-Python}  \footnote{\href{http://github.com/SciServer/SciScript-Python}{http://github.com/SciServer/SciScript-Python}} and {\em SciScript-R} \footnote{\href{http://github.com/SciServer/SciScript-R}{http://github.com/SciServer/SciScript-R}} wrap HTTP requests to the REST APIs of these components, passing along a user's token to ensure authentication (and authorization). The libraries can also be used remotely from a user's own machine. The SciServer components accessible via SciScript are the Login Portal, CasJobs, Files Service, Compute Jobs (see below), SciQuery, SkyServer, and SkyQuery. One common use case is users launching SQL queries to the databases in CasJobs, and getting the resulting tabular data set into a Python session as a pandas\footnote{\href{https://pandas.pydata.org}{https://pandas.pydata.org}} data frame for further analysis. This use case directly addresses the original requests from SDSS users for server side analysis and visualisation of SQL query results.

        \subsection{Batch Jobs}\label{Section:BatchJobsInCompute}
        Interactive Jupyter sessions are very convenient for exploring data sets and developing analysis pipe lines. However, the synchronous nature of these sessions makes them less suited for executing large, data intensive jobs. Also, the demands on compute power and memory do not fit well within the interactive compute domains, which may run up to a hundred containers on a single node concurrently. 
        Therefore an asynchronous, batch mode is required for these larger jobs.
        
        SciServer Compute supports such batch jobs in the form of executable Docker Containers. The jobs running in these containers can be plain shell commands, or entire Jupyter Notebooks executed using Jupyter's \texttt{NBconvert} command. 
        
        SciServer has a number of batch job compute domains available, distinguished by the type and size of available compute resources,
         and different configurations of concurrent usages and maximum job life time. Special examples of these are domains with GPUs supporting Docker images with CUDA\footnote{\href{https://developer.nvidia.com/cuda-zone}{https://developer.nvidia.com/cuda-zone}}, TensorFlow\footnote{\href{https://www.tensorflow.org/}{https://www.tensorflow.org/}} and PyTorch\footnote{\href{https://pytorch.org/}{https://pytorch.org/}} for executing machine learning pipelines.
        
        The SciServer Compute Jobs UI allows users to define jobs by choosing the requested domain, Docker image, the data and/or user volumes they need to be mounted, the notebook or shell command they want to have executed and a working folder where the job is to be executed. The job is submitted to the JOBM Rest API in RACM (see Figure \ref{FIG:Components}), which stores it in the Jobs queue in its database. The Docker Compute-Manager (COMPM) service sends requests to the JOBM API for the next job queued for execution. We have a custom solution to decide which job should be next, which takes into account a user's recent usage on the compute domain. Then COMPM requests the Compute app's REST API to launch a new executable Docker container within which the job is run and which is configured based on the job definition's metadata. COMPM follows the state of the jobs execution, which can be tracked also on the Compute UI's jobs page.
        Users can also use SciScript to programmatically submit jobs to Compute. The main use case for this is to avoid having to manually enter lots of job definitions for a parameterized workflow.

\section{SciServer Usage}\label{Section:Usage}

Here we describe how and by whom the SciServer instance at JHU is being used. 

\subsection{Usage statistics}
\begin{figure}
    	\centering
    		\includegraphics[scale=.50]{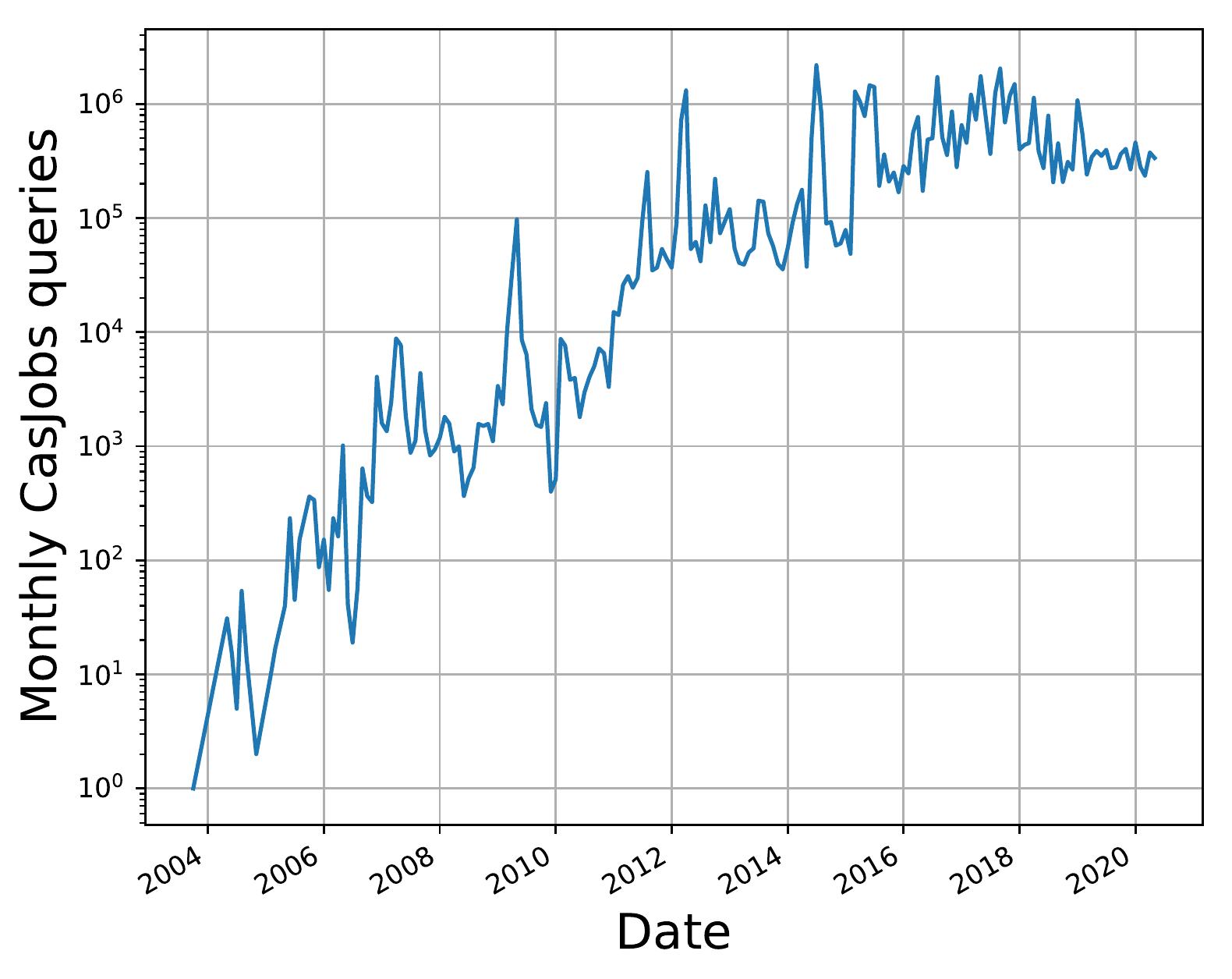}
    	\caption{Time evolution of the number of monthly CasJobs queries. Peaks in the figure correspond to releases of new database contexts, such as SDSS catalogs.}
    	\label{FIG:CasJobsQueries}
    \end{figure}

\begin{figure}
    	\centering
    		\includegraphics[scale=.50]{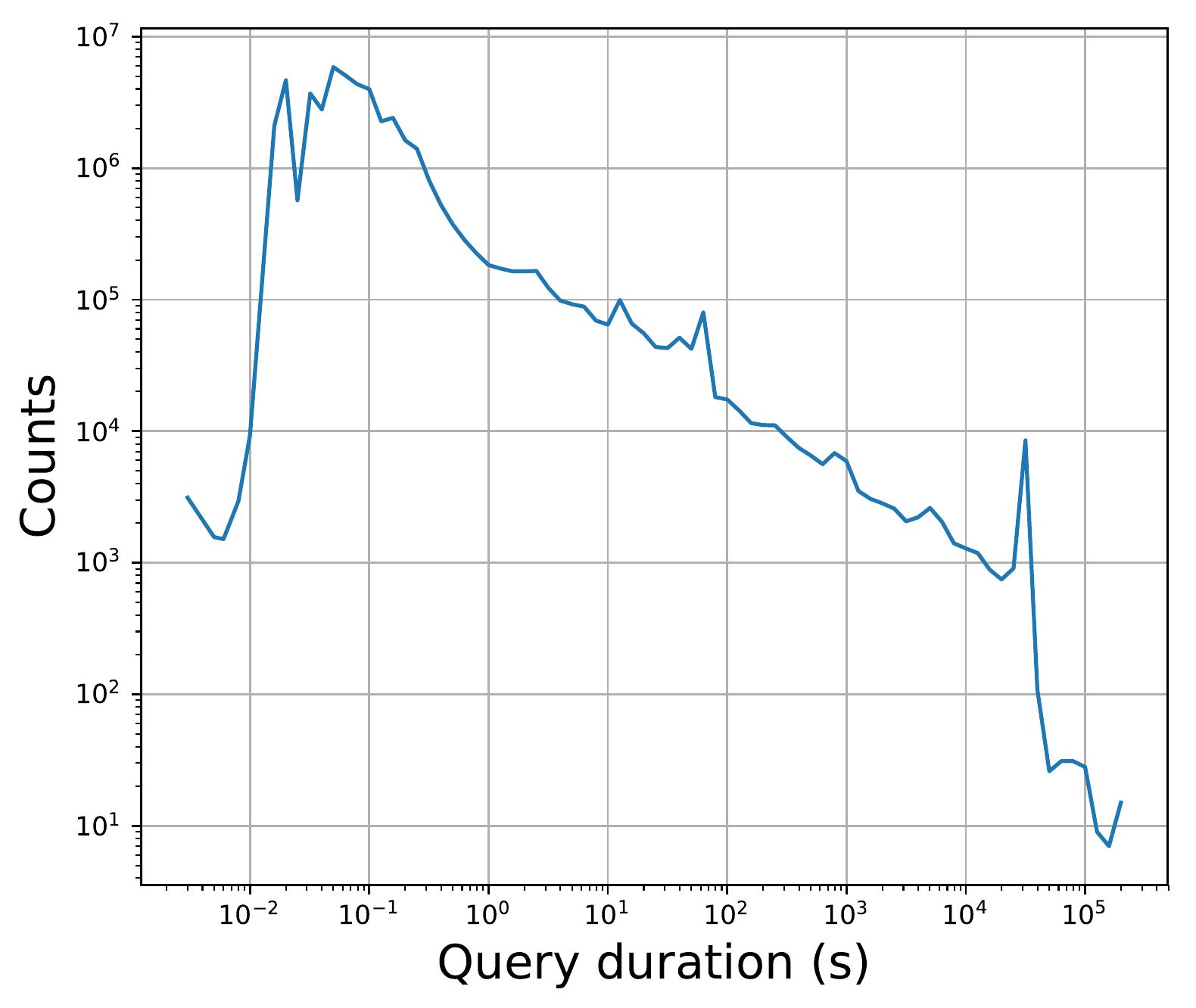}
    	\caption{Histogram of duration of CasJobs queries (in equal-size log-space bins). Note the power-law shape, with peaks at 1min and 8hrs, corresponding to the timeouts for synchronous and asynchronous queries, respectively.}
    	\label{FIG:CasJobsQueryTimeouts}
    \end{figure}

Usage data for SkyServer - the progenitor of SciServer - has been logged since the beginning (2001), and up-to-the-minute usage statistics are available on the SkyServer traffic page\footnote{\href{http://skyserver.sdss.org/log/en/traffic/}{http://skyserver.sdss.org/log/en/traffic/}} that can be accessed via the "Site Traffic" link on the SkyServer front page. Earliest records of queries in CasJobs go back to 2003, as seen in Fig. \ref{FIG:CasJobsQueries}. Both reflect the steady and widespread use of these resources by the worldwide astronomy community as well as the public. The number of queries increased sharply in the first few years and has stabilized over the last few years as most of the astronomy community has learned to use these tools. Peaks in the query submission rate are mainly driven by the publishing of new data, such as SDSS data releases. The distribution of query execution time is similar to a power-law (Fig. \ref{FIG:CasJobsQueryTimeouts}), with peaks at 1 min and 8 hrs, corresponding to the timeouts for synchronous and asynchronous queries, respectively, in CasJobs. 

\begin{figure}
    	\includegraphics[scale=.25]{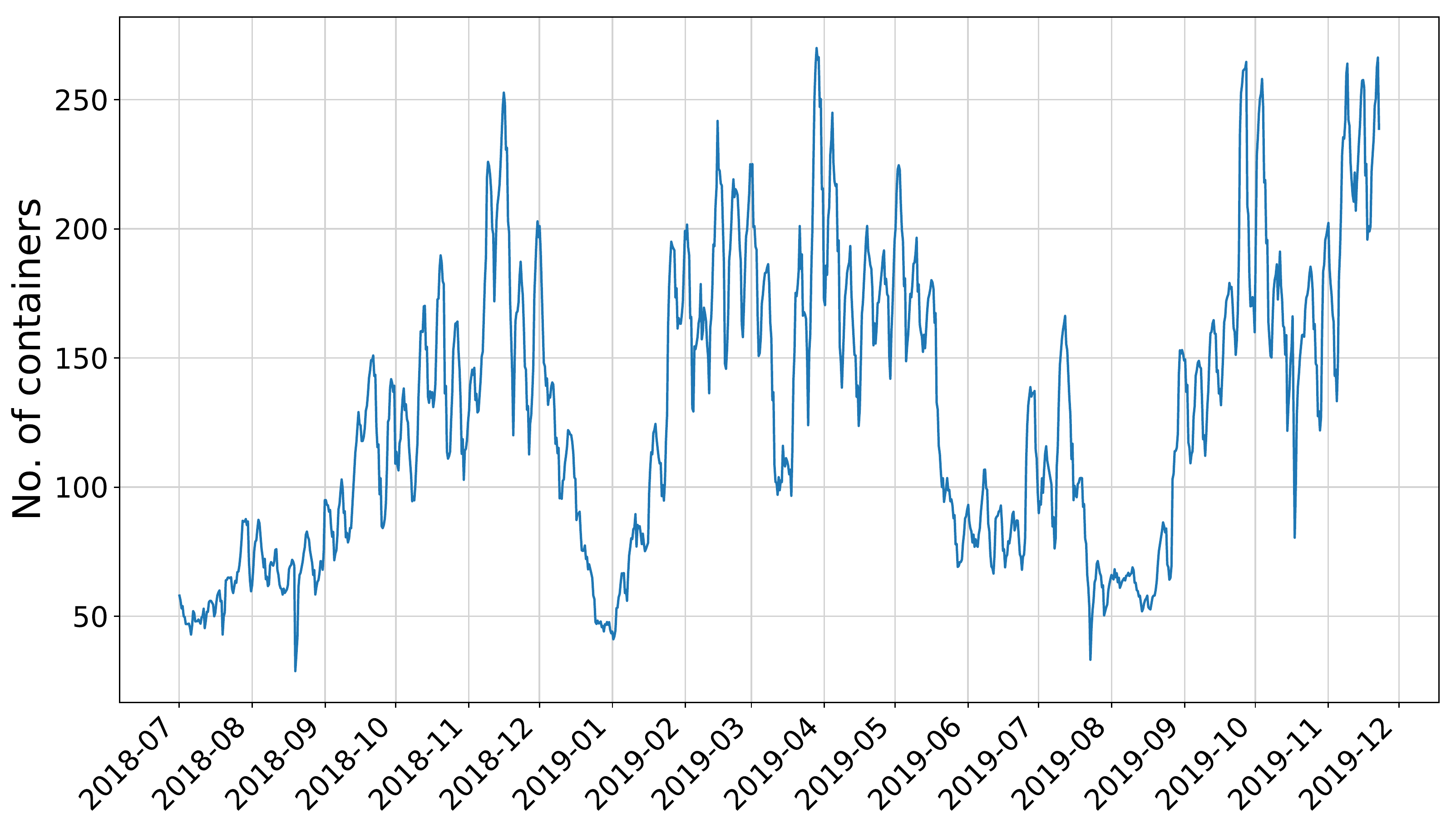}
    	\caption{Number of interactive Docker containers running in SciServer Compute at any given time since the July 2018 SciServer release. One can observe that activity decreases over weekends, and is overall driven by vacation/work schedules. This would seem to indicate that students are a significant part of its user base.}
    	\label{FIG:ActiveContainers}
    \end{figure}

\begin{figure}
    	\centering
    		\includegraphics[scale=.50]{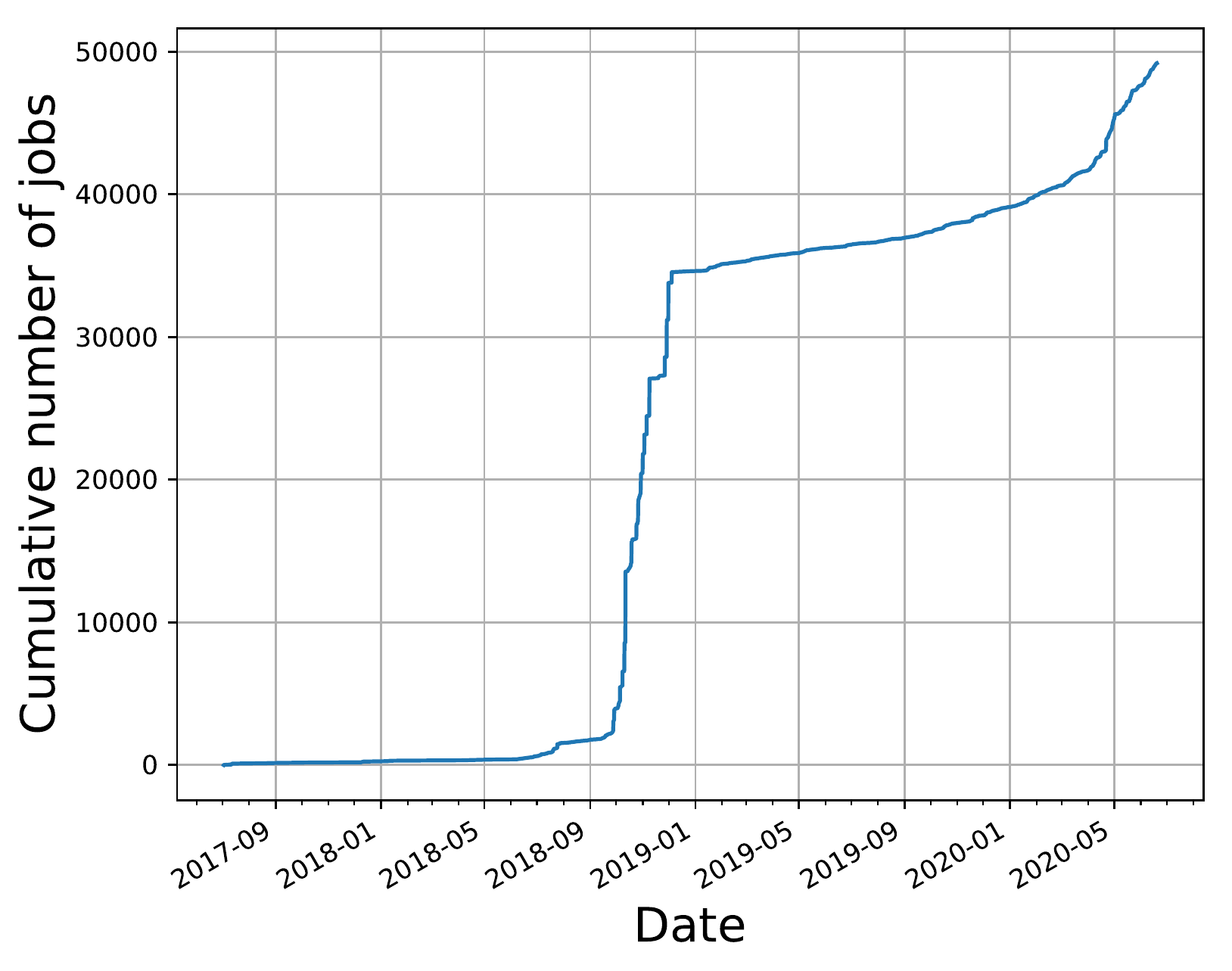}
    	\caption{Cumulative number of asynchronous jobs submitted to SciServer-Compute as a function of time. The big jump around November 2018 corresponds to users submitting programmatically multiple jobs by using the SciScript-Python client library during one specific SciServer project.}
    	\label{FIG:JobsEvolution}
    \end{figure}

With the advent of SciServer as a science platform, usage patterns have changed, as expected. SciServer Compute usage shows a substantial number of interactive sessions at any given time, as shown in Fig. \ref{FIG:ActiveContainers}. With a baseline of around 50, it has reached maximums of up to ~270 live concurrent sessions. One can observe the decrease in activity during weekends, and it appears to be mainly driven by the vacation/work schedules, which seems to indicate students are also a part of its user base.
The cumulative number of asynchronous jobs submitted to SciServer Compute is shown in Fig. \ref{FIG:JobsEvolution}. An example of the flexibility of SciServer APIs can be observed in the sharp increase of jobs in November 2018. On that occasion, users programmatically submitted multiple jobs by means of the SciScript-Python library for one specific SciServer project. Job submissions continue to increase significantly, both through the interactive website and SciScript client libraries.

\begin{figure}
    	\centering
    		\includegraphics[scale=.50]{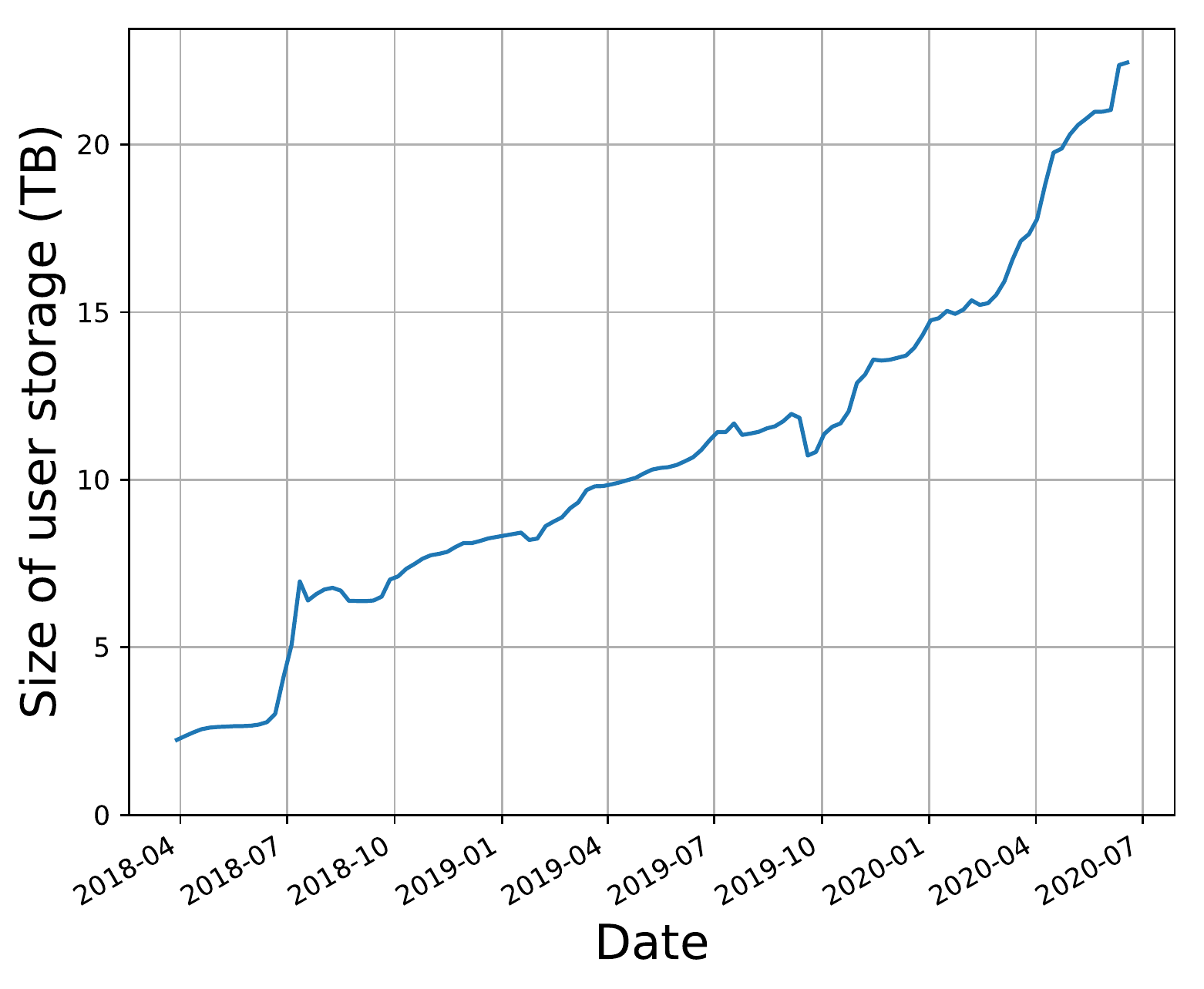}
    	\caption{Time evolution of total size of user data files in both Storage and Temporary volumes. Bulk of files are either uploaded by users, derived from bigger database tables or  files in public data volumes, or created during the data analysis process.}
    	\label{FIG:UserData}
    \end{figure}

Along with computing activity, the total size of user-created data files, both in the Storage and Temporary volumes in SciServer Compute, has risen to more than 20TB, as seen in Fig. \ref{FIG:UserData}. Total size is mostly dominated by either files that were uploaded by users, derived from bigger database tables or files in public data volumes, or created during the data analysis process.

    \subsection{Event logging and user history} \label{SubSection:EventLogging}

        Most of the user activity events are logged, and subsequently shown to the users in a searchable user history page within the SciServer Dashboard application. We only include meaningful events triggered by means of REST API calls or backend code, such as creation of new Docker containers, user volumes, file uploads and downloads, execution of SQL queries, user sign-ups and log-ins, creation of groups and sharing of resources, etc. System and REST API errors are also logged, including stack traces. Events occurring in SciServer apps are serialized into a JSON formatted message and temporarily stored in a RabbitMQ queue \footnote{\href{https://www.rabbitmq.com}{https://www.rabbitmq.com}}. Then, the Log Listener service consumes these messages from the queue and inserts them into the SciServerLog MS-SQL database. 
	MS-SQL supports the storage of JSON messages as columns with full search capabilities within the JSON message objects (see Figure \ref{FIG:Components}).
	Current work includes using ElasticSearch \footnote{\href{https://www.elastic.co}{https://www.elastic.co}} as a more flexible storage solution.
	In order to expose the contents of this event database, the SciServer Logging REST API can be invoked for searching, filtering and retrieving JSON formatted event messages that can be displayed in any browser or client.

\subsection{Science support} \label{SubSection:ScienceSupport}
One of the goals of extending the SkyServer and CasJobs applications to the SciServer platform was to support a larger set of scientific disciplines. IDIES had already been involved in building a database holding results of simulations of hydrodynamic turbulence and a web application for its dissemination to the public\footnote{\href{http://turbulence.pha.jhu.edu}{http://turbulence.pha.jhu.edu}} \citep{JHTDB2008}.
One of the authors built a database holding results of cosmological simulations at the MPA in Munich, Germany \citep{Lemson2006}. The database and its web application were directly inspired by the SDSS database and the CasJobs service, and it showed that astrophysicists were willing to learn SQL also for this type of data.

Both these data sets are now integrated in SciServer as well. The turbulence database is accessible through a special purpose Docker Image deployed in SciServer Compute, and the Millennium databases have been copied over to JHU and are accessible for SQL querying in the CasJobs tool.

\begin{figure*}
        	\centering
        		\includegraphics[scale=.11]{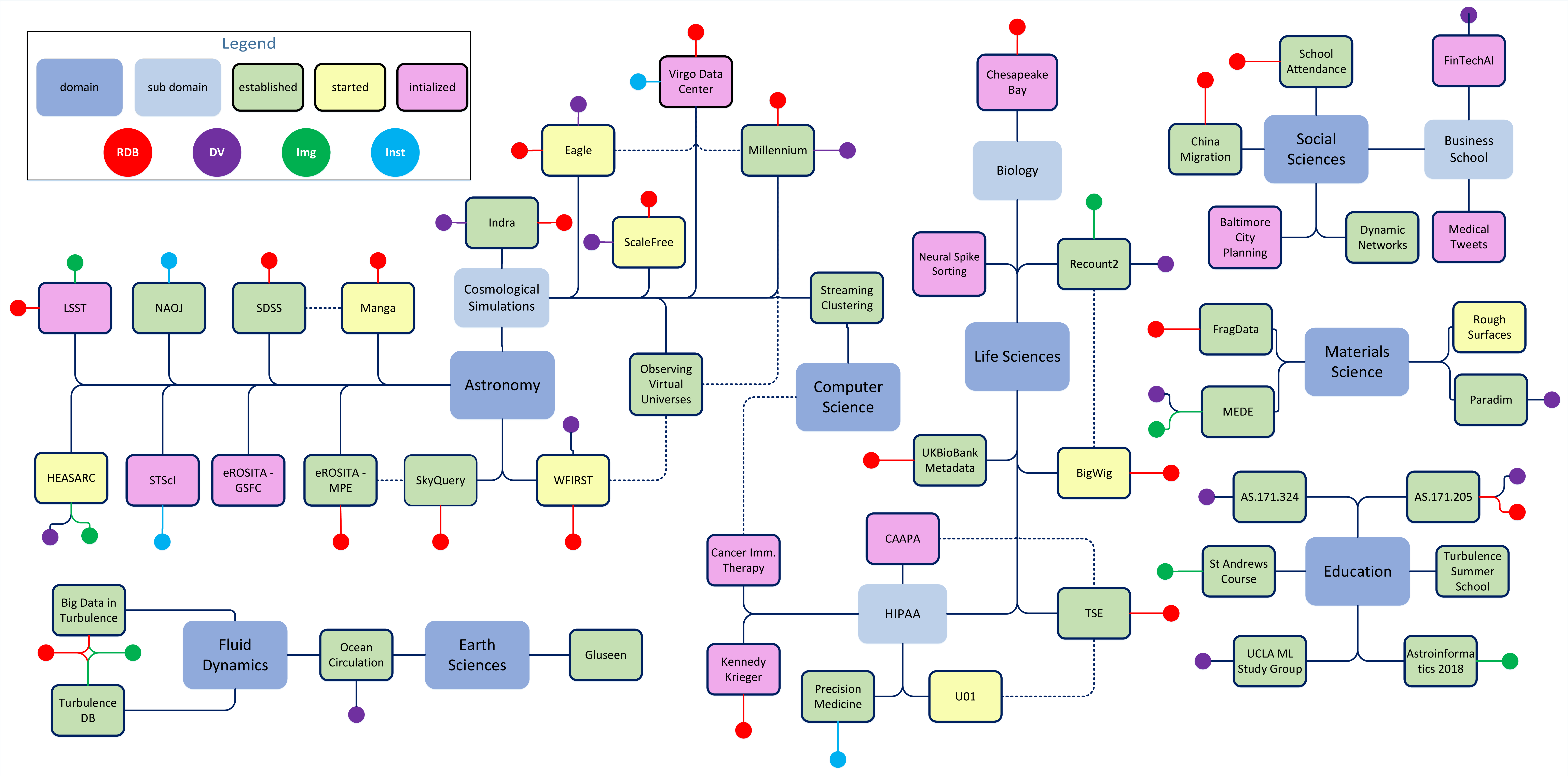}
        	\caption{This diagram illustrates the projects that are supported by SciServer at JHU and/or have served as use cases for its development. Though Astronomy remains a very important discipline for us, many more have been added, up to social and life sciences. The blue boxes indicate the domains, green boxes projects that are well embedded with SciServer. Yellow and pink boxes have either just started or have only had some initial contacts. The little circles indicate how these projects have been supported. This can be through the creation of custom databases, data volumes or compute images for the specific projects. Some special cases are those that got their own installation of SciServer, indicated by the light blue circle.}
        	\label{FIG:ScienceDomains}
        \end{figure*}

However, the development of SciServer was guided by many more scientific projects, as illustrated in the diagram in Figure~\ref{FIG:ScienceDomains}. The various cells in the diagram show projects that were science drivers for the development of SciServer.
The disciplines supported to date include life and social sciences, materials science and genomics. SciServer has also been used in classroom and schools.
Many of these projects use SciServer to host and provide access to data sets,  some for dissemination to the public, some for collaborative projects from a few colleagues to large consortia.
SciServer supports these disciplines sometimes with custom-built databases, published through Cas\-Jobs; sometimes by providing Terabytes of storage space and mounting those as data volumes; sometimes by building and providing Compute images with specialized soft\-ware required for the projects.

The inclusion of projects from many different science domains can lead to confusion for SciServer users. For example, the Millennium data set contains over twenty separate databases. When made public, these will all be visible in the various CasJobs selection menus and there be mixed in with the original SDSS databases. To allow people to select only those publicly accessible databases and other resources they are interested in, we have introduced the concept of a Science Domain. This is like a user group that users can freely join or leave, and to which particular resources have been added. For example, the "Astronomy" science domain will provide access to the SDSS databases and data volumes, the "Cosmology" domain will give access to the suite of Millennium, Eagle and Indra cosmological simulations. Similarly, there will be domains for Genomics, Oceanography and other areas for which our instance will host publicly accessible data sets.

These considerations all pertain only to the SciServer instance at JHU, where we support multiple science domains. In a few cases, we have provided separate standalone versions of SciServer, managed by the projects themselves on their premises. For example, the Max-Planck institute for extraterrestrial physics in Munich, Germany will deploy its own instance to support the eROSITA and HetDex projects. The National Astronomical Observatory of Japan (NAOJ) has an installation for supporting the SuMIRe project's HSC-PFS (Hyper Suprime Camera - Prime Focus Spectrograph) data set \citep{Takada2012}, where a subset of SciServer has been deployed on a Linux cluster using Kubernetes.
A special case is the Precision Medicine Analysis Platform built by the school of Public Health at JHU. They use a version of SciServer, named Crunchr, deployed inside a "safe desktop" \footnote{\href{http://ictr.johnshopkins.edu/programs\_resources/programs-resources/i2c/secure-research-data-desktop/}{http://ictr.johnshopkins.edu/programs\_resources/programs-resources/i2c/secure-research-data-desktop/}} that provides secure access conforming to the requirements on personalized health data.

    \subsection{SciUI}\label{SubSection:SciUI}
 SciServer gives its users an extremely flexible and powerful interface for accessing and analyzing its data sets, nevertheless some data providers prefer that their users are presented with more basic user interfaces. For example, though the turbulence databases described above could in principle be queried with SQL, the owners of the data preferred a simple interface, where users can enter a small set of parameters to retrieve a subset of the data. This was an acceptable solution as long as the data sets that were retrievable remained small, but this no longer scales when users request cutouts of the data that  extend to multiple Gigabytes, and the execution of which can no longer be handled using interactive requests.

We built a solution to this that uses SciServer as the backend implementation behind a simple, parameterized frontent UI. This solution serves as a template for other similar requests, and was coined \textbf{SciUI}.
        SciUI is a model for web applications or gateways that can be created external to SciServer, but that use the access controls, computing and data storage resources available in SciServer for simplifying the data analysis to a particular niche community or use case. 
        A SciUI application provides an easy-to-follow form-based interface for submitting particular jobs to SciServer, together with a view of the list of jobs and links for downloading the job results.
        In the turbulence example, the web application was written in Python Flask \footnote{\href{https://flask.palletsprojects.com}{https://flask.palletsprojects.com}} and uses the SciScript-Python library for logging in users and accessing data sets and job computing resources. After job submission and execution, the final results are written into a dedicated "jobs" user volume that is also accessible to the user from SciServer Compute if further analysis is required.
        Current examples of SciUI are the above mentioned Turbulence Cutout  \footnote{\href{http://turbulence.idies.jhu.edu/cutout}{http://turbulence.idies.jhu.edu/cutout} } for  creating cutouts from turbulence simulations, the Sequencer \footnote{\href{http://sequencer.org}{http://sequencer.org} } application for sorting and revealing trends in generic data sets, and the Tomographer \footnote{\href{http://tomographer.org}{http://tomographer.org} } application for computing the redshift tomography of astronomical source catalogs or intensity maps.

\subsection{Education use cases} \label{SubSection:Education}
        
        The fact that SciServer enables server-side analysis close to the data makes it ideal for use in science and technology education. Basic programming skills are an increasingly important part of workforce development, and students can learn these skills by solving real-world data analysis problems.
        
        With SciServer, students will no longer need to spend hours or days installing and customizing programming environments - they can begin doing meaningful science immediately. The fact that all students, instructors, and TAs are using the same platform removes problems due to software version incompatibilities, and ensures that everyone is using the same features. Furthermore, SciServer's collaboration features make it especially amenable to be used in a classroom setting where teachers need to selectively share course materials and resources with their students. Thus, teachers can share assignment folders (with read-only access) with students, students can share their completed assignments with (only) the teacher, etc.
        
        A number of instructors have already begun using SciServer to teach courses in astronomy and other subjects. These use cases range from using SciServer to offer lab activities in a large introductory astronomy class setting, to teaching a full semester-long lab course for third-year astronomy majors, conducted entirely in the SciServer environment. A number of training workshops for graduate students have similarly used the SciServer environment to enable quick-start research and education. Hands-on SciServer sessions have featured prominently in the "SDSS in the Classroom" workshops held at the American Astronomical Society (AAS) annual meetings in the past few years, and will do so again at the Honolulu AAS meeting in January 2020.
        
        Now that many instructors around the world are productively using SciServer for education, our next goal is to make it as easy as possible to manage instruction. 
        
        To help instructors create learning activities in the SciServer environment, and to help them manage their students' work through those activities, we are developing a new resource called \textit{SciServer Courseware}. Courseware will allow instructors to provide data and scripts to their students with appropriate permissions, and custom software environments in which to run those scripts. For its implementation, Courseware will make extensive use of the RACM and Files Service APIs.

\section{Summary and Future Work}
By reusing, re-engineering and extending the building blocks that SDSS created for server-side astronomical data analysis, we have developed SciServer - a full-featured, scalable, portable and interoperable science platform that facilitates collaborative data-driven science in astronomy and oth\-er sciences. 

The Jupyter-based Compute subsystem, along with a lay\-er of REST APIs to connect to the data warehouse, provides a powerful and versatile server-side analysis facility with access to several Petabytes of data (including all major astronomy data sets) via SQL or other access protocols, in interactive and batch modes. Collaborative science, both in a research and a classroom setting, is enabled by the Groups and access control features that we built in. A scripting environment makes it easy to execute complex tasks as workflows. 

Scalability and interoperability are built into SciServer with its Docker/VM architecture and a layer of REST APIs. Portability has been enhanced with the use of Docker/VM, the migration of all existing components to .NET Core, the use of Python and Java for all new components, the modular REST API based architecture, and the use of Kubernetes to deploy SciServer components. Portability and interoperability have been tested with SciServer deployments at other sites and running Docker images from other science platforms within SciServer (e.g. LSST, Montage). We have also test-deployed SciServer Compute successfully in the AWS cloud.

We will continue to work on enhancing the portability of SciServer by migrating services from CasJobs to the platform-agnostic SciQuery, and ultimately phasing CasJobs out. Future work for SciQuery will include synchronous query capability and query constructs that will allow users to freely create and share databases in either read-only or read/write mode, as is the case for the files in user volumes currently. We aim to separate the MyDB functionality from CasJobs and make it part of SciQuery in the future.

Another area of emphasis for SciServer in the near future will be expanding support for machine learning (ML) applications. SciServer has limited support for GPUs at the moment and has made available to users ML images that come packaged with ML libraries. Researchers have been using this for applications like classification of data using neural nets in TensorFlow.  

Courseware will continue to be developed as a  resource that can assist teaching in classrooms, and we expect to release it within the next year.

SciServer has received bridge funding to continue these activities from the National Science Foundation  via a supplement to the original DIBBs grant, and from the Moore Foundation.  Meanwhile, we are working towards building a self-sustainable funding model for SciServer as a service offered by JHU/IDIES.

Our long-term goal is to be able to deploy SciServer in all the major commercial cloud service platforms so that data providers will have many more options for how they will use SciServer. We expect to continue to work with other astronomy science platforms to enhance interoperability between the different platforms and ensure that there is a healthy exchange of ideas.

\section{Acknowledgements}
SciServer is funded by the United States National Science Foundation (award \#ACI-1261715, Data Infrastructure Building Blocks), the Sloan Foundation (award \#G-2018-11216) and the Gordon and Betty Moore Foundation (award \#8004). SciServer is supported and administered by the Institute for Data Intensive Engineering and Science (IDIES \footnote{\href{http://idies.jhu.edu/}{http://idies.jhu.edu}}) at the Johns Hopkins University. We are grateful to all SciServer team members\footnote{\href{http://www.SciServer.org/about/the-SciServer-team/}{http://www.SciServer.org/about/the-SciServer-team/}}, past and present, for their contributions to making SciServer such a success. We would also like to thank the SDSS collaboration for adopting SciServer as their official science platform for catalog data, and their ongoing contributions to improvements and enhancements of SciServer components. Special thanks to Rita Tojeiro and Anne-Marie Weijmans of the University of St Andrews, Karen Masters of Haverford College, Britt Lundgren of the University of North Carolina Asheville, and Michael Blanton of New York University for all their efforts in promoting use of SciServer in the classroom, and all their valuable feedback that helps us to hone its classroom features.

\appendix

\section{Resource Access Control Management: design and implementation}
\label{Appendix:RACM}
SciServer's support for collaboration and sharing is one of its unique features compared to other science platforms. It was described at a high level in Section~\ref{Section:RACM}, here we describe its design and implementation in greater detail.

\begin{figure*}
    	\centering
    		\includegraphics[scale=.18]{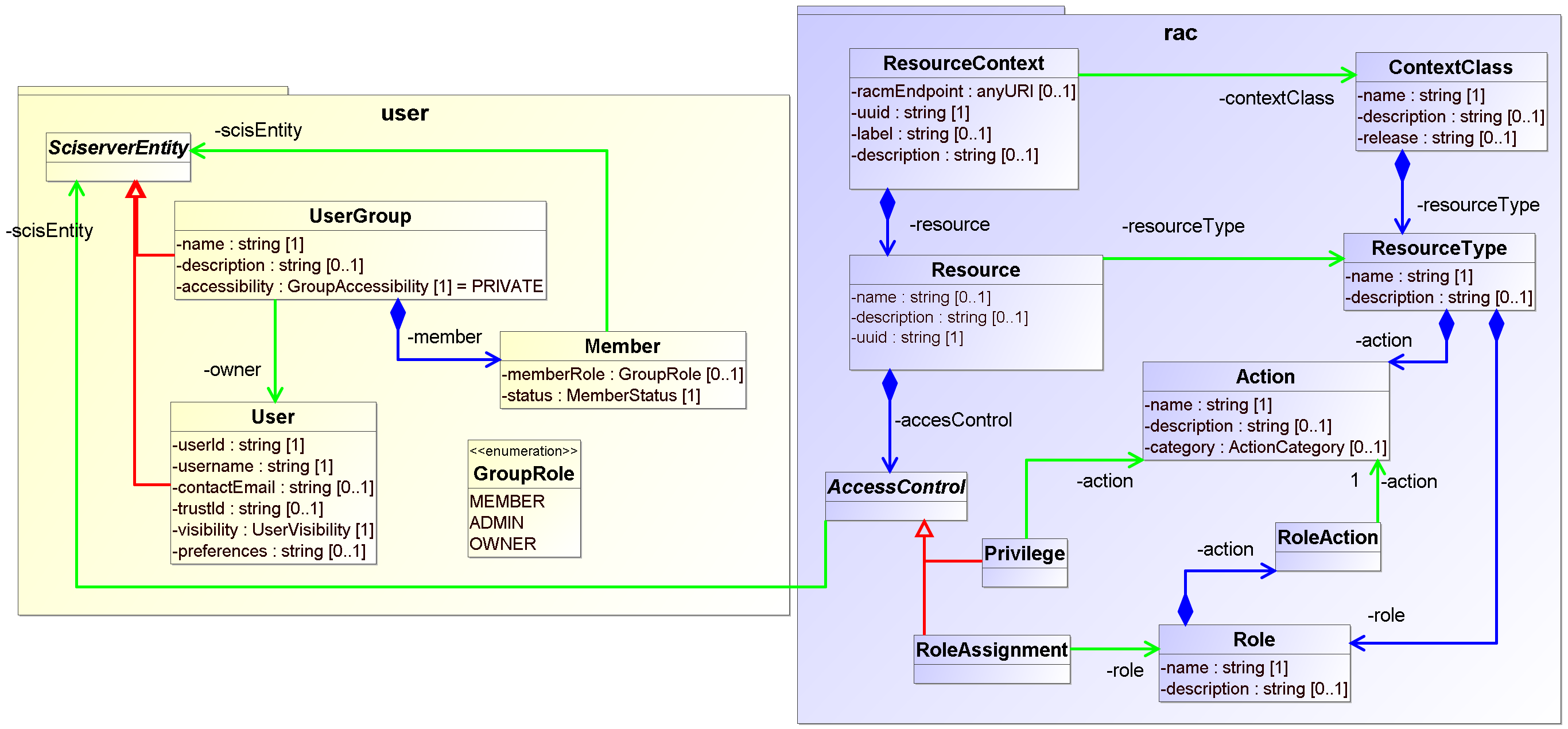}
    	\caption{UML version of the object model used in the resource access control management component in SciServer. Rectangles are classes, with attributes indicated. Red arrows indicate inheritance relations, blue lines are parent-child relationships and green arrows are references, implying a kind of shared usage relationship. Most of the classes are described in the Appendix.}
    	\label{FIG:ClassDiagramRACM}
    \end{figure*}

This Resource Access Control Management  component, RACM, is based on a formal UML object model illustrated in Figure~\ref{FIG:ClassDiagramRACM}.

    The model defines the important concepts as UML classes with various types of relationships defining inheritance, composition and references. Central concept is the \textbf{Resource}. This represents the entities that can be shared between users and groups in SciServer. 
    In  the model, Resources are owned by a \textbf{ResourceContext}, as represented by the (blue) parent-child composition relation. These  ResourceContext instances are represented in the "real" world by (web) applications such as a Files Service, or the CasJobs database access tool. The Resources they own, manage, correspond to shareable entities such as databases, data sets or compute environments.
    
    Each Resource references (the green arrow) a  \textbf{ResourceType}. RACM has a number of predefined ResourceTypes in its registry. Examples are "Database", "DataVolume", "DockerImage" etc, and they define the type of the Resource that is being shared. The ResourceType also defines \textbf{Actions} that can be "executed" on Resource-s of that type. Examples are "read" and "write" on a "UserVolume", or "submitQuery" on a "Database".
    When sharing a Resource, its owner should indicate which Action(s) the sharee is given the privilege to execute. This is implemented in the model 
    by assigning \textbf{Privileges} to a \textbf{User} or \textbf{UserGroup}. User and UserGroup are also represented as a class and note that the structure of the model implies that groups can be nested. 

RACM only stores metadata about shared resources, but is otherwise oblivious to their meaning. The implication of a Privilege assignment is completely implemented by the application represented by the ResourceContext. 
As an example, take the CasJobs web application at JHU. It hosts some databases that are not accessible to all users. To support this in RACM,  CasJobs is first represented as a ResourceContext. This context can define resource of  ResourceType {\em DatabaseContext}, which has an  Action  named {\em submitQuery}.

When a new database is added to CasJobs, it will be represented by such a Resource and to give access to this database to a user or group, a Privilege is created on the new Resource, linking the User or UserGroup and the action "submitQuery".
When a user connects to CasJobs, it queries RACM for all the database contexts it manages \textit{and that the user has access to}, either directly or because they are member of a group that was assigned the privilege. This listing will be used to populate metadata components on the web application's UI. But it will also be used to check that the user has appropriate privileges on each of the databases they try to access when submitting a SQL query.

    A special Action is "grant". Regardless of resource type, all users with \textit{grant} privileges on a Resource can give access permissions for that Resource to other users or groups. This is a generalization of the ownership of a resource.
        
The RACM object model is implemented using an updated version of the VO-URP\footnote{\href{https://github.com/glemson/vo-urp}{https://github.com/glemson/vo-urp}} tool developed originally in support of the Simulation Data Model in the IVOA \citep{Lemsonetal2012}. This tool generates Java code and a relational database model from the object model specified in UML\footnote{\href{https://www.uml.org}{https://www.uml.org}}. VO-URP comes with infrastructure code that manages the storage and retrieval of Java objects to/from the database using an object-relational mapping prescription implemented on the Java classes using annotations following the Java Persistence Architecture\footnote{\href{https://en.wikipedia.org/wiki/Java_Persistence_API}{https://en.wikipedia.org/wiki/Java\_Persistence\_API}}.
    RACM wraps this code with APIs and is deployed as a standalone web application.The other SciServer components such as the Dashboard use these APIs, and so do some of the SciScript libraries defined in Section~\ref{SubSection:SciScript} when they need to interact directly with the RACM backend.
    
    Currently, RACM is dependent on the MS-SQL database for its implementation, though the free MS-SQL Express edition is sufficient for most purposes. A Postgres implementation of VO-URP used to exist, but has not been kept up to date; remedying this situation is one of our short-term goals.
        
\bibliographystyle{cas-model2-names}

\bibliography{sciserver}

\end{document}